\documentclass[12pt]{article}
\usepackage[round]{natbib} 

\usepackage{amssymb}
\usepackage{amsmath}
\usepackage{MnSymbol}
\usepackage{graphicx}
\usepackage{float}
\usepackage{ragged2e}
\usepackage{setspace}
\usepackage{blindtext}
\usepackage{bbm}
\usepackage{latexsym}
\usepackage{caption}
\usepackage{hyperref}
\usepackage{float}
\usepackage{comment}
\usepackage{dcolumn}
\usepackage{adjustbox}
\usepackage{xcolor}
\usepackage{setspace}
\usepackage{authblk}
\usepackage{caption}
\usepackage{subcaption}

\newcommand{\blind}{0}

\newcommand\independent{\protect\mathpalette{\protect\independenT}{\perp}}
\def\independenT#1#2{\mathrel{\rlap{$#1#2$}\mkern2mu{#1#2}}}

\begin{document}

\def\spacingset#1{\renewcommand{\baselinestretch}%
{#1}\small\normalsize} \spacingset{1}


\if0\blind
{
  \title{\bf A Nonparametric Bayesian Methodology for Regression Discontinuity Designs}
  \author{Zach Branson\thanks{
    This research was supported by the National Science Foundation Graduate Research Fellowship Program under Grant No. 1144152, by the National Science Foundation under Grant No. 1461435, by DARPA under Grant No. FA8750-14-2-0117, by ARO under Grant No. W911NF- 15-1-0172, and by NSERC. Any opinions, findings, and conclusions or recommendations expressed in this material are those of the authors and do not necessarily reflect the views of the National Science Foundation, DARPA, ARO, or NSERC.}\hspace{.2cm}\\
    \vspace{-0.1 in}
    Department of Statistics, Harvard University \\

    Maxime Rischard \\
    Department of Statistics, Harvard University \\

    Luke Bornn \\
    Department of Statistics and Actuarial Science, Simon Fraser University 

    Luke Miratrix \\
    Graduate School of Education, Harvard University}
  \maketitle
} \fi

\if1\blind
{
  \bigskip
  \bigskip
  \bigskip
  \begin{center}
    {\LARGE\bf Title}
\end{center}
  \medskip
} \fi

\vspace{-0.35 in}

\begin{abstract}
One of the most popular methodologies for estimating the average treatment effect at the threshold in a regression discontinuity design is local linear regression (LLR), which places larger weight on units closer to the threshold. We propose a Gaussian process regression methodology that acts as a Bayesian analog to LLR for regression discontinuity designs. Our methodology provides a flexible fit for treatment and control responses by placing a general prior on the mean response functions. Furthermore, unlike LLR, our methodology can incorporate uncertainty in how units are weighted when estimating the treatment effect. We prove our method is consistent in estimating the average treatment effect at the threshold. Furthermore, we find via simulation that our method exhibits promising coverage, interval length, and mean squared error properties compared to standard LLR and state-of-the-art LLR methodologies. Finally, we explore the performance of our method on a real-world example by studying the impact of being a first-round draft pick on the performance and playing time of basketball players in the National Basketball Association.

\end{abstract}

\noindent%
\vfill

\newpage
\spacingset{1.45} 

\section{Introduction} \label{s:introduction}

Recently there has been a renewed interest in regression discontinuity designs (RDDs), which originated with \cite{thistlethwaite1960regression}. In an RDD, the treatment assignment is discontinuous at a certain covariate value, or ``threshold,'' such that only units whose covariate is above the threshold will receive treatment. There are many examples of RDDs in the applied econometrics literature: the United States providing additional funding to only the 300 poorest counties for the Head Start education program \citep*{ludwig2007does}; schools mandating students to attend summer school if their exam scores are below a threshold \citep*{matsudaira2008mandatory}; colleges offering financial aid to students whose academic ability is above a cutoff \citep*{van2002estimating}; and Medicare increasing insurance coverage after age 65 \citep*{card2004impact}. The main goal of an RDD is to estimate a treatment effect while addressing likely confounding by the covariate that determines treatment assignment. 

One of the most popular methodologies for estimating the average treatment effect at the threshold in an RDD is local linear regression (LLR), which places larger weight on units closer to the threshold. Implementation of LLR is straightforward and there is a wide literature on its theoretical properties. However, recent works have found that LLR can exhibit poor inferential properties---such as confidence intervals that tend to undercover---which has motivated a strand of literature started by \cite{calonico2014robust} that modifies LLR to improve coverage and other inferential properties.

Adding to this literature, we propose a nonparametric regression approach that acts as a Bayesian analog to LLR for sharp regression discontinuity designs. Our approach utilizes Gaussian process regression (GPR) to provide a flexible fit for treatment and control responses by placing a general prior on the mean response functions. While GPR has been widely used in the machine learning and statistics literature, it has not previously been proposed for estimating treatment effects in RDDs. Thus, our main contribution is outlining how to use Gaussian processes to make causal inferences in RDDs and assess how such a methodology compares to current LLR methodologies.

In the remainder of this section, we review RDDs and LLR methodologies for estimating the average treatment effect at the threshold. In Section \ref{s:gpMethod}, we outline GPR for sharp RDDs and note various analogies to LLR, which builds intuition for implementing our method. In Section \ref{s:consistency}, we establish that our method is consistent in estimating the average treatment effect at the boundary. In Section \ref{s:simulations}, we show via simulation that our method exhibits promising coverage, interval length, and mean squared error properties compared to standard LLR and state-of-the-art LLR methodologies. In Section \ref{s:empiricalExample}, we use GPR on data from the National Basketball Association (NBA) to estimate the effect of being a first-round versus a second-round pick on basketball player performance and playing time, and we find that GPR detects treatment effects that are more in line with previous results in the sports literature than do LLR methodologies. In Section \ref{s:conclusion}, we conclude by discussing extensions to our methodology to tackle problems beyond sharp RDDs.

\subsection{Overview of Regression Discontinuity Designs}

We follow the notation of \cite{imbens2008regression} and discuss RDDs within the potential outcomes framework: For each unit $i$, there are two potential outcomes, $Y_i(1)$ and $Y_i(0)$, corresponding to treatment and control, respectively, and a covariate $X_i$. Only one of these two potential outcomes is observed, but $X_i$ is always observed. Let $W_i$ denote the assignment for unit $i$, where $W_i = 1$ if $i$ is assigned to treatment and $W_i = 0$ if unit $i$ is assigned to control. The observed outcome for unit $i$ is then
\begin{align}
  y_i = W_i Y_i(1) + (1 - W_i)Y_i(0)
\end{align}
Often, one wants to estimate the average treatment effect $\mathbb{E}[Y_i(1) - Y_i(0)]$, but usually this treatment effect is confounded by $X$ (and possibly other unobserved covariates), such that examining the difference in mean response between treatment and control is not appropriate. In these cases, methods such as stratification, matching, and regression are often employed to address covariate confounding when estimating the average treatment effect. However, such methods are only appropriate when there is sufficient overlap in the treatment and control covariate distributions, i.e.,
\begin{align}
  0 < P(W_i = 1 | X_i = x) < 1 \label{overlapAssumption}
\end{align}
For example, this overlap assumption is essential for propensity score methodologies, where the relationship between $W_i$ and $X_i$ is estimated and then accounted for during the analysis. 

In an RDD, the relationship between the treatment assignment $W_i$ and the covariates is known. More specifically, it is known that the function $P(W = 1 | X = x)$ is discontinuous at some threshold or boundary $X = b$. In this paper we focus on a special case, \textit{sharp RDDs}, where the treatment assignment for a unit $i$ is
\begin{align}
  W_i = \begin{cases}
    1 &\mbox{if } X_i \geq b \\
    0 &\mbox{otherwise.}
  \end{cases} \label{sharpAssignment}
\end{align}
The covariate $X$ that determines treatment assignment in an RDD is often called the ``running variable'' in order to not confuse it with other background covariates that do not necessarily determine treatment assignment. The more general case, where $P(W = 1 | X = x)$ is discontinuous at $X = b$ but is not necessarily 0 or 1, is known as a \textit{fuzzy RDD}. 

\cite{rubin1977assignment} states that when treatment assignment depends on one covariate, estimating $\mathbb{E}[Y_i(1) | X_i]$ and $\mathbb{E}[Y_i(0) | X_i]$ is essential for estimating the average treatment effect. Furthermore, treatment effect estimates are particularly sensitive to model specification of $\mathbb{E}[Y_i(1) | X_i]$ and $\mathbb{E}[Y_i(0) | X_i]$ when there is not substantial covariate overlap, as in a sharp RDD. Aware of this sensitivity, RDD analyses typically do not attempt to estimate the average treatment effect. Instead, they focus on the estimand that requires the least amount of extrapolation to overcome this lack of covariate overlap: the average treatment effect \textit{at the boundary} $b$. Defining the conditional expectations for treatment and control as
\begin{align}
  \mu_T(x) \equiv \mathbb{E}[Y_i(1) | X_i = x], \hspace{0.1 in} \mu_C(x) \equiv \mathbb{E}[Y_i(0) | X_i = x] ,
\end{align}
the treatment effect at the boundary $b$ is (Imbens and Lemieux 2008)
\begin{align}
  \tau = \lim_{x \downarrow b} \mathbb{E}[y_i | X_i = x] - \lim_{x \uparrow b} \mathbb{E}[y_i | X_i = x] = \mu_T(b) - \mu_C(b) . \label{tau}
\end{align}
The notation $\mu_T(x)$ and $\mu_C(x)$ emphasizes that the goal of an RDD requires estimating two unknown mean functions. \cite{hahn2001identification} showed that sufficient conditions for $\tau$ to be identifiable in a sharp RDD are that $\mu_C(x)$ and $\mu_T(x) - \mu_C(x)$ are continuous at $b$. They further state that ``we can use any nonparametric estimator to estimate'' $\mu_T(x)$ and $\mu_C(x)$, and recommended local linear regression (hereafter called LLR), which is currently the most popular methodology for estimating the average treatment effect at the boundary in an RDD.

\subsection{Review of Local Linear Regression}

The goal of an RDD is to estimate $\tau$ defined in (\ref{tau}), i.e., to estimate $\mu_T(b)$ and $\mu_C(b)$. LLR estimates $\mu_T(b)$ as
\begin{align}
  \hat{\mu}_T(b) = X_b (X_T^T W_T X_T)^{-1} X_T^T W_T Y_T \label{LLREstimator}
\end{align}
where $X_b \equiv \begin{pmatrix} 1 & b \end{pmatrix}$, $X_T$ is the $n_T \times 2$ design matrix corresponding to the intercept and running variable $X$ for treated units, $Y_T$ is the $n_T$-dimensional column vector of treated units' responses, and $W_T$ is a $n_T \times n_T$ diagonal weight matrix whose entries are
\begin{align}
  (W_T)_{ii} \equiv K \hspace{-0.025 in} \left( \frac{x_i - b}{h} \right), \hspace{0.1 in} i = 1, \dots, n_T \label{weightMatrix}
\end{align}
for some kernel $K(\cdot)$ and bandwidth $h$. The estimator $\hat{\mu}_C(b)$ is analogously defined for the control. Then, the estimator for the treatment effect is $\hat{\tau} = \hat{\mu}_T(b) - \hat{\mu}_C(b)$. Here, $\hat{\mu}_T(b)$ and $\hat{\mu}_C(b)$ are weighted least squares estimators, where the weights depend on units' closeness to the boundary $b$. To perform local polynomial regression, one appends higher orders of $X$ to the design matrices $X_T$ and $X_C$.

The RDD literature has focused on LLR largely because of its boundary bias properties. For example, \cite{hahn2001identification} recommend LLR over alternatives like kernel regression because \cite{fan} showed that LLR exhibits better bias properties for boundary points than kernel regression. For more details on the bias comparison between kernel regression and LLR, see \citealt{imbens2008regression} (p. 624-625) as well as \cite{fanAndGijbels} and \cite{porter2003estimation}. 

Furthermore, LLR's implementation is straightforward once a kernel $K(\cdot)$ and bandwidth $h$ are chosen. The most common choice of $K(\cdot)$ is the rectangular or triangular kernel; \cite{imbens2008regression} argue that more complicated kernels rarely make a difference in estimation. Much more attention has been given to the bandwidth choice $h$, largely because the bias is characterized by $h$. In the 2000s, choosing an appropriate $h$ for LLR in RDDs was an open problem: For example, \cite{ludwig2007does} stated that ``there is currently no widely-agreed-upon method for selection of optimal bandwidths...so our strategy is to present results for a broad range of candidate bandwidths.'' One widely-used bandwidth selection method is that of \cite{imbens2012optimal}, who derived a data-driven, MSE-optimal bandwidth for LLR estimators. This provided practitioners with clear guidelines for implementing LLR for RDDs, which made its use very popular.

The bandwidth is arguably the most important choice to be made in the LLR methodology for RDDs, because the treatment effect is often sensitive to the bandwidth choice. This motivates sensitivity checks such as that in \cite{ludwig2007does}, where the treatment effect is estimated several times with different bandwidths to ensure that estimates do not vary too greatly. Some have noted that confidence intervals from LLR have a tendency to undercover when a single bandwidth is chosen for inference when the treatment effect is sensitive to the bandwidth choice \citep{armstrong2017simple,gelman2018high}. A recent extension to the LLR methodology---that of \cite{calonico2014robust}, hereafter called ``robust LLR''---was one of the first methods to address the undercoverage issue of LLR by incorporating a bias correction and inflated confidence intervals corresponding to the uncertainty in estimating the bias correction. Because of its promising inference properties, \cite{calonico2014robust} has arguably become the state-of-the-art for conducting inference for the average treatment effect at the boundary in an RDD.

\subsection{Other Methods Besides LLR}

Other methodologies besides LLR have been proposed for estimating the average treatment effect at the boundary in a sharp RDD. For example, many practitioners have used high-order global polynomials to estimate $\mu_T(x)$ and $\mu_C(x)$: \cite{matsudaira2008mandatory} argued for a global third-order polynomial regression, and also considered fourth- and fifth-order polynomials as a sensitivity check; similarly, \cite{van2002estimating} used a global third-order polynomial and noted that LLR could have been an alternative; finally, \cite{card2004impact} argued for using a global third-order polynomial regression instead of LLR because the running variable, age, was discrete. However, in recent years many have argued against the use of high-order polynomials in RDDs because of their tendency to yield point estimates and confidence intervals that are highly sensitive to the order of the polynomial \citep{calonico2015optimal,gelman2018high}.

Others have focused on local randomization methodologies, where units within a window around the boundary are viewed as-if randomized to treatment and control. For example, \cite{cattaneo2015randomization} recommends a series of covariate balance tests to decide the window around the boundary such that the as-if randomized assumption is most plausible. \cite{li2015evaluating} extended these ideas to develop a notion of a local overlap assumption and used a Bayesian hierarchical modeling approach for deciding the window around the boundary where this assumption is most plausible. \cite{cattaneo2017comparing} compared the local randomization approach to local polynomial estimators, and they extended the local randomization approach to incorporate adjustments via parametric models as well.

Finally, others have developed Bayesian methodologies for RDDs. \cite{li2015evaluating} propose a principal stratification approach that provides alternative identification assumptions based on a formal definition of local randomization. \cite{geneletti} propose a Bayesian methodology that incorporates prior information in the treatment effect. \cite{chib2014nonparametric} use Bayesian splines to estimate treatment effects in RDDs. \cite{chib2015} propose a Bayesian methodology specific to fuzzy RDDs.

\subsection{Our Proposal: Gaussian Process Regression for Sharp RDDs}

We propose a methodology that utilizes Gaussian process regression (GPR), which is one of the most popular nonparametric methodologies in the machine learning and Bayesian modeling literature for estimating unknown functions \citep{rasmussenWilliams}. The notion of using GPR for RDDs is very much in line with the claim in \cite{hahn2001identification} that any nonparametric estimator can be used to estimate the treatment and control response in an RDD. However, to our knowledge, GPR has not been previously proposed for RDDs.

Similar to LLR, our GPR methodology provides a flexible fit to the mean functions $\mu_T(x)$ and $\mu_C(x)$. Furthermore, our methodology can incorporate both prior knowledge and uncertainty in various parameters in the RDD problem---such as how units are weighted near the boundary---which is not necessarily as straightforward with current LLR methodologies. Finally, our GPR methodology can be used in conjunction with a local randomization perspective. Our methodology adds to the strand of literature started by \cite{calonico2014robust} that addresses the undercoverage of standard LLR, as well as the strand of literature on Bayesian methodologies for RDDs.

\section{GPR Models to Estimate the Average Treatment Effect at the Boundary} \label{s:gpMethod}

First we review notation for GPR and how GPR is used to estimate a single unknown function. We then discuss GPR models that estimate the two unknown mean functions in sharp RDDs. 

\subsection{Notation for Estimating One Unknown Function} \label{ss:gprReview}

Define a dataset $\{x_i, y_i\}_{i=1}^n$ of responses $\mathbf{y} = (y_1, \dots, y_n)$ that varies around some unknown function of the covariate $\mathbf{x} = (x_1, \dots, x_n)$:
\begin{align}
  \mathbf{y} = f(\mathbf{x}) + \epsilon
\end{align}
where $\epsilon \sim \mathcal{N}_n(0, \sigma^2_y \mathbf{I}_n)$ and $f(\mathbf{x}) \equiv (f(x_1),\dots,f(x_n))$. If the goal is to well-estimate $\mathbb{E}[f(x^*)]$ for a particular covariate value $x^*$, one option is to specify a functional form for $f(\mathbf{x})$ and then predict $\mathbb{E}[f(x^*)]$ from this specified model, such as local linear regression, as discussed in Section \ref{s:introduction}. Instead of specifying a functional form for $f(\mathbf{x})$, we consider nonparametrically inferring $f(\mathbf{x})$ by placing a prior on $f(\mathbf{x})$. A Gaussian process is one such prior:
\begin{align}
  f(\mathbf{x}) \sim \text{GP}(m(\mathbf{x}), K(\mathbf{x}, \mathbf{x}'))
\end{align}
for some unknown mean function $m(\mathbf{x})$ and covariance function $K(\mathbf{x}, \mathbf{x}')$. The notation $f(\mathbf{x}) \sim GP(m (\mathbf{x}), K(\mathbf{x}, \mathbf{x}'))$ denotes a Gaussian process prior on the unknown function $f(\mathbf{x})$, which states that, for any $(x_1, \dots, x_n)$, the joint distribution $(f(x_1), \dots, f(x_n))$ is an $n$-dimensional multivariate normal distribution with mean vector $(m(x_1), \dots, m(x_n))$ and covariance matrix $K(\mathbf{x}, \mathbf{x}')$ whose $(i,j)$ entries are $K(x_i, x_j')$. 

There are many choices one could make for the mean and covariance functions. A common choice for the mean function is $m(\mathbf{x}) = \mathbf{0}$; a common choice for the covariance function is the squared exponential covariance function, whose entries are
\begin{align}
  K(x_i, x_j) \equiv \sigma^2_{GP} \exp \left(- \frac{1}{2 \ell^2} (x_i - x_j)^2 \right). \label{squaredExponential}
\end{align}
Placing a Gaussian process prior with a squared exponential covariance function on $f(\mathbf{x})$ assumes that $f(\mathbf{x})$ is infinitely differentiable, which is similar to other assumptions in the RDD literature (e.g., Assumption 3.3 of \citealt{imbens2012optimal} and Assumption 1 of \citealt{calonico2014robust}). The covariance parameters $\sigma^2_{GP}$ and $\ell$ are called the variance and lengthscale, respectively. The variance determines the amplitude of $f(\mathbf{x})$, i.e., how much the function varies from its mean. The lengthscale determines the smoothness of the function: Small lengthscales correspond to $f(\mathbf{x})$ changing rapidly. Most importantly, the covariance function assumes that the response at a particular covariate value $f(x^*)$ will be similar to the response at covariate values close to $x^*$.

Given the Gaussian process prior with mean function $m(\mathbf{x})$ and covariance function $K(\mathbf{x}, \mathbf{x}')$, as well as their parameters, the posterior for $f(x^*)$ at any particular covariate value $x^*$ can be obtained via standard conditional multivariate normal theory (for an exposition, see \citealt{rasmussenWilliams}, Pages 16-17):
\begin{equation}
\begin{aligned}
  f(x^*) | \mathbf{x}, \mathbf{y} &\sim N(\mu^*, \sigma^2_{GP} - \Sigma^*), \hspace{0.1 in} \text{where}
\end{aligned}
\end{equation}
\begin{equation}
\begin{aligned}
  \mu^* &\equiv m(x^*) + K(x^*, \mathbf{x})[K(\mathbf{x}, \mathbf{x}) + \sigma^2_y \mathbf{I}_n]^{-1} (\mathbf{y} - m(\mathbf{x})) \\
  \Sigma^* &\equiv K(x^*, \mathbf{x})[K(\mathbf{x}, \mathbf{x}) + \sigma^2_y \mathbf{I}_n]^{-1} K(\mathbf{x}, x^*)
\end{aligned}
\end{equation}
The above posterior can thus be used to obtain point estimates and credible intervals for the value of an unknown function at a particular covariate value $x^*$. In practice, the covariance parameters are estimated from the data, such as through maximum likelihood or cross-validation \citealt{rasmussenWilliams} (Chapter 5). In Section \ref{ss:gpModels} we assume that the covariance parameters are fixed, and in Section \ref{ss:covarianceUncertainty} we extend to a full Bayesian approach that places priors on $\ell, \sigma^2_{GP}$, and $\sigma^2_y$.

\subsection{GPR Models for Sharp RDDs} \label{ss:gpModels}

The notion of using GPR to estimate the average treatment effect at the boundary in an RDD suggests a class of models that has not previously been considered in the RDD literature. We focus on two GPR models, which correspond to different assumptions placed on the unknown response functions $\mu_T(x)$ and $\mu_C(x)$. For each model we show the resulting posterior for the average treatment effect at the boundary and compare it to its analogous LLR model. In Section \ref{ss:covarianceUncertainty} we discuss how---unlike LLR methodologies---the uncertainty in how units are weighted can be incorporated into these GPR models. 

For both GPR models, we assume that the treatment response $Y_i(1)$ and the control response $Y_i(0)$ have the following relationship with the running variable $x_i$
\begin{equation}
\begin{aligned}
    &Y_i(1) = \mu_T(x_i) + \epsilon_{i1}, \hspace{0.1 in} \text{and} \hspace{0.1 in} Y_i(0) = \mu_C(x_i) + \epsilon_{i0}, \hspace{0.1 in} \text{where} \\
  &\mu_T(x_i) \independent \mu_C(x_i), \hspace{0.1 in} \epsilon_{i1} \stackrel{iid}{\sim} N(0, \sigma_{y1}^2), \hspace{0.1 in} \text{and } \epsilon_{i0} \stackrel{iid}{\sim} N(0, \sigma_{y0}^2) \label{gpModelEquation}
\end{aligned}
\end{equation}
Thus, intuitively, the procedure outlined in Section \ref{ss:gprReview} can simply be performed twice---once for $\mu_T(x)$ and once for $\mu_C(x)$. However, there are assumptions on $\mu_T(x)$ and $\mu_C(x)$ that, if true, can simplify our GPR models and make inference more precise.

In particular, assumptions can be placed on the covariance structure of $\mu_T(x)$ and $\mu_C(x)$. The two models we present correspond to two different sets of assumptions---the first assumes that the covariance structure of $\mu_T(x)$ and $\mu_C(x)$ are the same, while the second allows them to be different. In both models, we assume that $\mu_T(x)$ and $\mu_C(x)$ are stationary processes, i.e., the covariance parameters of $\mu_T(x)$ and $\mu_C(x)$ do not vary with the running variable $X$. We discuss cases when this stationarity assumption is inappropriate in Sections \ref{s:simulations} and \ref{s:conclusion}. \\

\noindent
\textbf{Same Covariance Assumption}: $\text{Cov}(\mu_T(x)) = \text{Cov}(\mu_C(x))$, and $\mu_T(x)$ and $\mu_C(x)$ are stationary processes. \\ 

If the Same Covariance Assumption holds, a natural LLR procedure is to fit local linear regressions on both sides of the boundary with the same bandwidth but different intercepts and slopes. This is largely the standard practice in the RDD literature \citep*{imbens2008regression}. Analogously, we place Gaussian process priors on $\mu_T(x)$ and $\mu_C(x)$ for given mean functions $m_T(\mathbf{x})$ and $m_C(\mathbf{x})$ and covariance function $K(\mathbf{x}, \mathbf{x}')$:
\begin{equation}
\begin{aligned}
    \mu_T(x) &\sim \text{GP}(m_T(\mathbf{x}), K(\mathbf{x}, \mathbf{x}')) \\
    \mu_C(x) &\sim \text{GP}(m_C(\mathbf{x}), K(\mathbf{x}, \mathbf{x}')) \label{twoGPMeanFunctions}
\end{aligned}
\end{equation}
Then, estimates $\hat{\mu}_T(b)$ and $\hat{\mu}_C(b)$ are obtained, which results in a treatment effect estimate $\hat{\tau} = \hat{\mu}_T(b) - \hat{\mu}_C(b)$. We now outline how such estimates $\hat{\mu}_T(b)$ and $\hat{\mu}_C(b)$ are obtained. Using standard conditional multivariate normal theory as in Section \ref{ss:gprReview}, we have the following posteriors for $\mu_T(b)$ and $\mu_C(b)$:
\begin{equation}
\begin{aligned}
  \mu_T(b) | \mathbf{x}, \mathbf{y} &\sim N(\mu_{b | T}, \sigma^2_{GP} - \Sigma_{b | T}) \\
  \mu_C(b) | \mathbf{x}, \mathbf{y} &\sim N(\mu_{b | C}, \sigma^2_{GP} - \Sigma_{b | C}), \hspace{0.1 in} \text{where}
\end{aligned}
\end{equation}
\begin{equation}
\begin{aligned}
  \mu_{b | T} &\equiv m_T(b) + K(b, \mathbf{x}_T)[K(\mathbf{x}_T, \mathbf{x}_T) + \sigma^2_y \mathbf{I}]^{-1} (\mathbf{y}_T - m_T(\mathbf{x}_T)) \\
  \mu_{b | C} &\equiv m_C(b) + K(b, \mathbf{x}_C)[K(\mathbf{x}_C, \mathbf{x}_C) + \sigma^2_y \mathbf{I}]^{-1} (\mathbf{y}_C - m_C(\mathbf{x}_C)) \\
  \Sigma_{b | T} &\equiv K(b, \mathbf{x}_T)[K(\mathbf{x}_T, \mathbf{x}_T) + \sigma^2_y \mathbf{I}]^{-1} K(\mathbf{x}_T, b) \\
  \Sigma_{b | C} &\equiv K(b, \mathbf{x}_C)[K(\mathbf{x}_C, \mathbf{x}_C) + \sigma^2_y \mathbf{I}]^{-1}K(\mathbf{x}_C, b) \label{covParamsMeanVar}
\end{aligned}
\end{equation}
Here, $\mu_{b | T}$ and $\mu_{b | C}$ denote the posterior mean for $\mu_T(b)$ and $\mu_C(b)$, respectively, which are in the definition of the treatment effect $\tau$ defined in (\ref{tau}). Note that $\mu_{b | T}$ and $\mu_{b | C}$ are weighted averages of the observed response, where the weights $K(b, \mathbf{x})[K(\mathbf{x}, \mathbf{x}) + \sigma^2_y \mathbf{I}]^{-1}$ depend on the covariance parameters $\ell$, $\sigma^2_{GP}$, and $\sigma^2_y$, as well as $\mathbf{x}$. For more discussion on the behavior of the weights $K(b, \mathbf{x})[K(\mathbf{x}, \mathbf{x}) + \sigma^2_y \mathbf{I}]^{-1}$, see \citet[Section 2.6]{rasmussenWilliams}. 

The posterior for the treatment effect under the Same Covariance Assumption is then
\begin{align}
  \tau \equiv \mu_T(b) - \mu_C(b) | \mathbf{x}, \mathbf{y} \sim N \left(\mu_{b | T} - \mu_{b | C}, 2 \sigma^2_{GP} - \Sigma_{b | T} - \Sigma_{b | C} \right). \label{posteriorTreatmentEffect}
\end{align}
where we have also used the independence of $\mu_T(x)$ and $\mu_C(x)$ stated in (\ref{gpModelEquation}). If the Same Covariance Assumption does not hold, one can still assume that the mean treatment and control response processes are stationary, but allow both the mean and covariance to vary on either side of the boundary. \\

\noindent
\textbf{Stationary Assumption}: $\mu_T(x)$ and $\mu_C(x)$ are stationary processes. \\ 

The posterior in this case would be identical to (\ref{posteriorTreatmentEffect}), except the means $\mu_{b | T}$ and $\mu_{b | C}$ and covariances $\Sigma_{b | T}$ and $\Sigma_{b | C}$ are instead defined as
\begin{equation}
\begin{aligned}
  \mu_{b | T} &\equiv m_T(b) + K_T(b, \mathbf{x}_T)[K_T(\mathbf{x}_T, \mathbf{x}_T) + \sigma^2_{y1} \mathbf{I}]^{-1} (\mathbf{y}_T - m_T(\mathbf{x}_T)) \\
  \mu_{b | C} &\equiv m_C(b) + K_C(b, \mathbf{x}_C)[K_C(\mathbf{x}_C, \mathbf{x}_C) + \sigma^2_{y0} \mathbf{I}]^{-1} (\mathbf{y}_C - m_C(\mathbf{x}_C)), \\
  \Sigma_{b | T} &\equiv K_T(b, \mathbf{x}_T)[K_T(\mathbf{x}_T, \mathbf{x}_T) + \sigma^2_{y1} \mathbf{I}]^{-1} K_T(\mathbf{x}_T, b) \\
  \Sigma_{b | C} &\equiv K_C(b, \mathbf{x}_C)[K_C(\mathbf{x}_C, \mathbf{x}_C) + \sigma^2_{y0} \mathbf{I}]^{-1}K_C(\mathbf{x}_C, b) \label{diffCovParamsMeanVar}
\end{aligned}
\end{equation}
i.e., the shared covariance $K(\cdot, \cdot)$ is replaced with $K_T(\cdot, \cdot)$ for units receiving treatment and $K_C(\cdot, \cdot)$ for units receiving control. The analogous LLR methodology would be to allow different intercepts, slopes, and bandwidths on either side of the boundary. However, using different bandwidths on either side of the boundary is rarely done in practice. For example, \cite{imbens2008regression} argue that if the curvature of $\mu_T(x)$ and $\mu_C(x)$ is the same, then the large-sample optimal bandwidths should be the same; and, furthermore, there is additional variance in estimating two optimal bandwidths rather than one, due to the smaller sample used to estimate each bandwidth. Thus, a benefit of the Same Covariance Assumption is that it allows researchers to use the entire data to estimate one set of covariance parameters, instead of estimating two separate sets of covariance parameters for treatment and control. However, when fitting our GPR model, we do not recommend sharing information between $\mu_T(x)$ and $\mu_C(x)$ beyond estimating their covariance structure---this follows the general practice in the RDD literature to fit separate regression functions (that may nonetheless share the same bandwidth) for the treatment and control groups.

The above posteriors for these two GPR models assume fixed mean and covariance parameters. In practice, maximum-likelihood or cross-validation can be used for estimating these parameters. In Section \ref{ss:covarianceUncertainty}, we extend the above GPR models to a full-Bayesian approach that incorporates uncertainty in the mean and covariance parameters.

\subsection{Accounting for Mean and Covariance Function Uncertainty} \label{ss:covarianceUncertainty}

The GPR models in Section \ref{ss:gpModels} assume that $\mu_T(b)$ and $\mu_C(b)$ will be similar to $\mu_T(x)$ and $\mu_C(x)$, respectively, for $x$ near $b$. The extent of this similarity is determined by the covariance function $K(\mathbf{x}, \mathbf{x}')$ and its parameters. In particular, recall that in Section \ref{ss:gpModels} we showed that the posterior mean of the average treatment effect for GPR is characterized by a difference of two weighted averages, where the weights are of the form $K(b, \mathbf{x})[K(\mathbf{x}, \mathbf{x}) + \sigma^2_y \mathbf{I}]^{-1}$. Thus, incorporating uncertainty in the covariance parameters in turn incorporates uncertainty in how units are weighted when estimating the average treatment effect. 

Denote the mean function parameters by $\boldsymbol{\theta}_m$ and the covariance function parameters by $\boldsymbol{\theta}_K$. For example, consider the mean functions
\begin{equation}
\begin{aligned}
  m_T(\mathbf{x}) &= \mathbf{h}(x)^T \boldsymbol{\beta}_T \\
  m_C(\mathbf{x}) &= \mathbf{h}(x)^T \boldsymbol{\beta}_C \label{linearMeanFunction}
\end{aligned}
\end{equation}
where $\mathbf{h}(x) = (1, x, \dots, x^{p-1})$, and $\boldsymbol{\beta}_T$ and $\boldsymbol{\beta}_C$ are the corresponding $p$-dimensional column vectors. In this case, $\boldsymbol{\theta}_m = (\boldsymbol{\beta}_T, \boldsymbol{\beta}_C)$. For the squared exponential covariance function defined in (\ref{squaredExponential}), $\boldsymbol{\theta}_K = (\ell, \sigma^2_{GP}, \sigma^2_y)$. 

In order to incorporate uncertainty in $\boldsymbol{\theta_m}$ and $\boldsymbol{\theta_K}$, one can first obtain draws $1, \dots, D$ from the joint posterior of $(\boldsymbol{\theta}_m, \boldsymbol{\theta}_K)$, rather than obtaining point-estimates $\hat{\boldsymbol{\theta}}_m$ and $\hat{\boldsymbol{\theta}}_K$. Then, for each draw $(\boldsymbol{\theta}_m, \boldsymbol{\theta}_K)_1, \dots, (\boldsymbol{\theta}_m, \boldsymbol{\theta}_K)_D$, one draws from the posterior for $\tau$, defined in (\ref{posteriorTreatmentEffect}). 

Section \ref{ss:gpModels} already defines the likelihood for the GPR models, so all that remains is to specify priors for $(\boldsymbol{\theta}_m, \boldsymbol{\theta}_K)$ in order to obtain draws from the joint posterior of $(\boldsymbol{\theta}_m, \boldsymbol{\theta}_K)$. Priors for $(\boldsymbol{\theta}_m, \boldsymbol{\theta}_K)$ will depend on the choice of mean and covariance functions. For example, for the mean functions defined in (\ref{linearMeanFunction}), we recommend $\mathcal{N}(\mathbf{0}, B)$ as a prior for $\boldsymbol{\beta}_T$ and $\boldsymbol{\beta}_C$, where $B$ is a $p \times p$ diagonal matrix with reasonably large entries (\citealt{rasmussenWilliams}, Pages 28-29). For the squared exponential covariance function given by (\ref{squaredExponential}), we recommend half-Cauchy priors for the covariance parameters $\frac{1}{\ell^2}$ and $\sigma^2_{GP}$ and noise $\sigma^2_y$, following advice from \cite{gelman2006} about stable priors for variance parameters. 

Now we prove that our GPR methodology is consistent in estimating the average treatment effect at the boundary. First we establish consistency when the covariance parameters are fixed, and then we consider the case where priors are placed on the covariance parameters.

\section{Posterior Consistency of our GPR Models} \label{s:consistency}

Gaussian processes are known to exhibit posterior consistency under minimal assumptions. \cite{ghosal2006posterior} proved posterior consistency of binary GPR for fixed covariance parameters, and \cite{choiAndSchervish} proved posterior consistency of GPR when the response is continuous. More generally, \cite{van2008rates} studied the contraction rate for Gaussian process priors for density estimation and regression problems, and \cite{van2009adaptive} extended these results to when a prior is placed on the lengthscale of a Gaussian process.

Here we evaluate our Bayesian methodology from a frequentist point-of-view, which assumes a fixed treatment effect at the boundary. The GPR models in Section \ref{ss:gpModels} estimate the treatment effect as the difference between two Gaussian process regressions; thus, our posterior of the treatment effect is consistent if the separate GPRs on either side of the discontinuity are consistent. First we establish posterior consistency assuming the covariance parameters are fixed, as in Section \ref{ss:gpModels}, and then we extend these results to when a prior is placed on the covariance parameters, as in Section \ref{ss:covarianceUncertainty}. 

We prove posterior consistency assuming the Stationary Assumption in Section \ref{ss:gpModels}, but the results also hold for the Same Covariance Assumption. Furthermore, we assume the mean functions $m_T(x) = 0$, $m_C(x) = 0$ and the squared exponential covariance function defined in (\ref{squaredExponential}). Discussion about the extent to which these results extend to other choices for the mean and covariance functions is in the Appendix. The other assumptions necessary for Theorems 1 and 2 below follow \cite{van2009adaptive} and are also given in the Appendix. \\

\noindent
\textbf{Theorem 1}: \textit{Assume that the Stationary Assumption holds, the covariance functions $K_T(x, x)$ and $K_C(x, x)$ are fixed, and Assumptions A1, A2, and A3 given in the Appendix hold. Denote the true average treatment effect at the boundary as $\tau^* = \mu_T^*(b) - \mu_C^*(b)$, where $\mu_T^*(x)$ and $\mu_C^*(x)$ are the true mean treatment and control response functions in the model (\ref{gpModelEquation}). Let $\prod(\tau | x_1, \dots, x_n)$ denote the posterior distribution of the average treatment effect at the boundary, defined in (\ref{posteriorTreatmentEffect}). Then, this posterior is consistent, in the sense that}
\begin{align}
  \prod \left( \tau: h(\tau, \tau^*) \geq M \epsilon_n | x_1, \dots, x_n \right) \xrightarrow{P_{\tau^*}} 0
\end{align}
\textit{for sufficiently large $M$, where $h$ is the Hellinger distance, and $\epsilon_{n}$ is the rate at which the posterior of $\tau$ contracts to the true $\tau^*$.} \\

\noindent
The proof of Theorem 1, as well as a discussion about the nature of the contraction rate, is given in the Appendix. Theorem 2 establishes posterior consistency when a prior is placed on the lengthscale parameter $\ell$, instead of being held fixed (see Section \ref{ss:covarianceUncertainty}). A discussion about posterior consistency when an additional prior is placed on $\sigma^2_{GP}$ is in the Appendix. \\

\noindent
\textbf{Theorem 2}: \textit{Assume that the Stationary Assumption holds, the $\sigma^2_{GP}$ parameters in $K_T(x, x)$ and $K_C(x, x)$ are fixed, and Assumptions A1, A2, A3, and A4 given in the Appendix hold. Then, Theorem 1 holds.} \\

\noindent
The proof of Theorem 2 is given in the Appendix. A corollary follows from the proofs of Theorem 1 and 2. \\

\noindent
\textbf{Corollary 1}: \textit{Theorems 1 and 2 hold if the Same Covariance Assumption holds instead of the Stationary Assumption.}

\section{Simulation Results} \label{s:simulations}

Here we investigate how our Gaussian Process model compares to standard LLR and the robust LLR method introduced in \cite{calonico2014robust}. We choose these two methods because the former is the standard in both applied work and the RDD literature at large, and the latter is a recent method that attempts to solve the undercoverage issue of standard LLR. We focus on the GPR model assuming the Same Covariance Assumption in Section \ref{ss:gpModels} and use the mean functions
\begin{equation}
\begin{aligned}
  m_T(\mathbf{x}) &= \boldsymbol{\beta}_{0T} + \boldsymbol{\beta}_{1T} \mathbf{x} \\
  m_C(\mathbf{x}) &= \boldsymbol{\beta}_{0C} + \boldsymbol{\beta}_{1C} \mathbf{x} \label{eqn:simulationMeanFunctions}
\end{aligned}
\end{equation}
and the squared exponential covariance function given by (\ref{squaredExponential}). These assumptions are analogous to the LLR procedure of fitting separate local linear regressions in treatment and control with differing slopes but the same bandwidth. Specification of the mean function in the Gaussian process prior is typically not consequential for estimation; however, as discussed in \citet[Section 2.7]{rasmussenWilliams}, there can be some benefits to specifying a mean function, as we do here. In particular, the above specification allows GPR predictions to pull towards a global linear trend instead of a global mean (which would be the case if we used a zero mean function---a common choice in the literature---for the Gaussian process prior). This can be useful within the context of extrapolation towards the boundary, as in an RDD. In the Appendix in Table \ref{tab:closeToBoundarySims}, we present simulation results for our GPR methodology using a zero mean function instead of the above linear mean function for the Gaussian process prior. The results for that case are largely the same as the results presented here, which suggests that our results are insensitive to specification of the mean function in the Gaussian process prior.

As discussed in Section \ref{ss:covarianceUncertainty}, we took a full-Bayesian approach to our GPR methodology and placed independent $\mathcal{N}(0,100^2)$ priors on the mean function parameters in (\ref{eqn:simulationMeanFunctions}) and independent $\text{half-Cauchy}(0, 5)$ priors on the covariance parameters. These choices for the prior distributions are in line with common recommendations in the Bayesian data analysis literature: The choice of Normal priors on the mean function parameters follows recommendations from \cite{rasmussenWilliams}, and the choice of half-Cauchy priors on the covariance parameters follows recommendations from \cite{gelman2006}, \cite{polson2012half}, and \citet[Chapter 5]{gelman2013bayesian}. We used the \texttt{R} package \texttt{rstan} \citep{carpenter2016stan} to sample from the posterior of these parameters. In the Appendix in Table \ref{tab:closeToBoundarySims}, we discuss simulation results for GPR when we instead plug in the MLE for the covariance parameters; however, we found that the full-Bayesian approach is preferable in terms of inferential properties, which suggests that it is beneficial to incorporate uncertainty in the covariance parameters for our GPR method.

We conduct a simulation study based on simulations from \cite{imbens2012optimal} and \cite{calonico2014robust}. In all simulations, we generated 1,000 datasets of 500 observations $\{(x_i, y_i, \epsilon_i): i = 1, \dots, 500 \}$, where $x_i \sim 2 \text{Beta}(2, 4) - 1$, $\epsilon_i \sim N(0, 0.1295^2)$, and $y_i = \mu_j(x_i) + \epsilon_i$ for different mean functions $\mu_j(x_i)$.

\begin{figure}[H]
\centering
  \includegraphics[scale = 0.7]{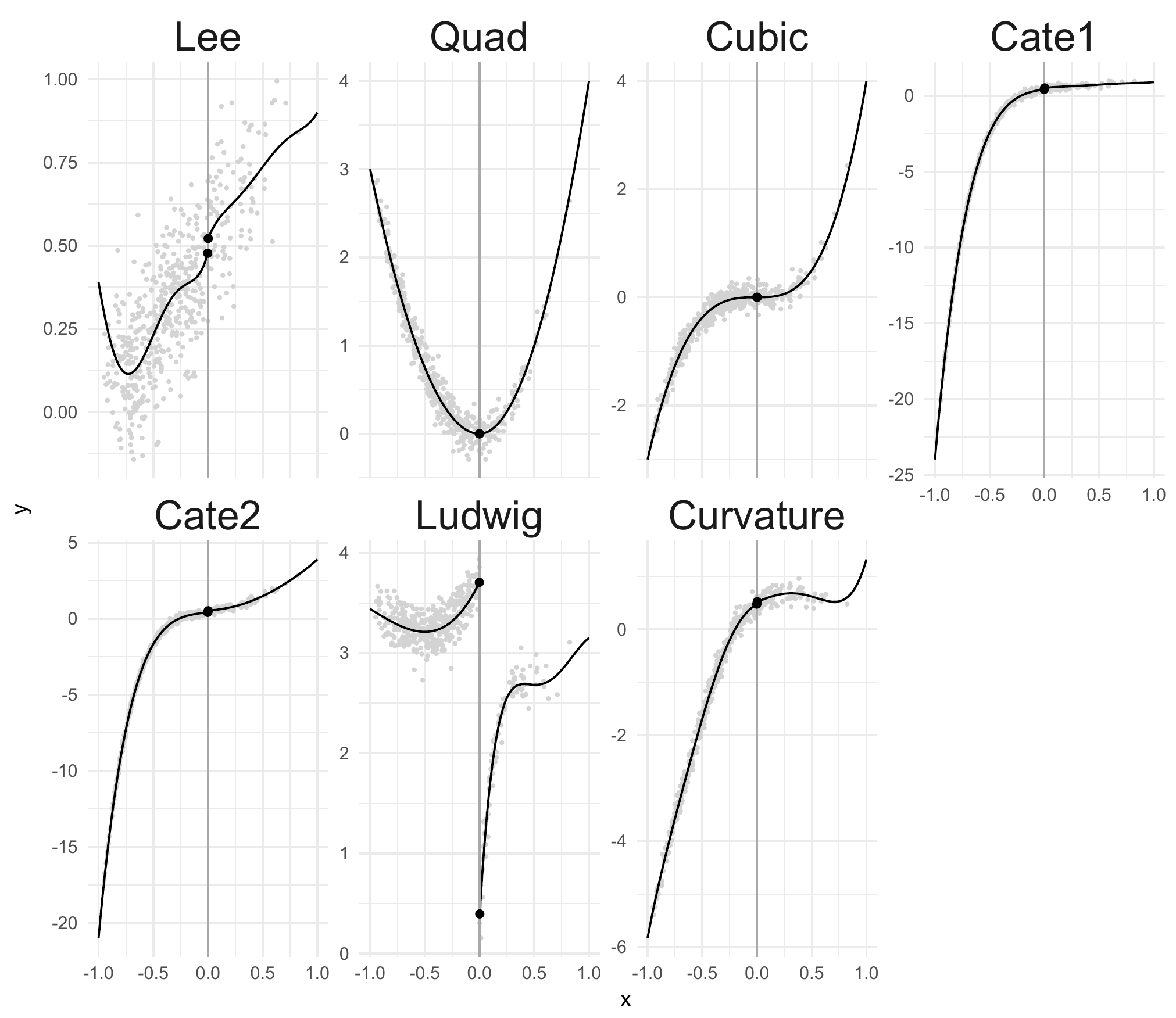}
  \caption{Mean function for the datasets used in our simulation study. Lee, Quad, Cate 1, and Cate 2 were used in \cite{imbens2012optimal}, and Lee, Ludwig, and Curvature were used in \cite{calonico2014robust}. Displayed in gray are a set of sample points $\{ (x_i, y_i), i = 1, \dots, 500 \}$ for each mean function.}
  \label{fig:meanFunctions}
\end{figure}

We consider seven different mean functions (see Figure \ref{fig:meanFunctions}), which we call Lee, Quad, Cate 1, Cate 2, Ludwig, Curvature, and Cubic. Lee, Quad, Cate 1, and Cate 2 were used in \cite{imbens2012optimal}, and Lee, Ludwig, and Curvature were used in \cite{calonico2014robust}; details about these datasets can be found in \citealt{imbens2012optimal} (Page 18) and \citealt{calonico2014robust} (Page 20). We also introduce the Cubic mean function as a comparison to the Quad mean function, because in the Quad mean function the linear trends on either side of $b$ are the opposite sign, whereas those for the Cubic mean function are the same sign. 

The boundary for each dataset is $b = 0$. The treatment effect is $\tau = 0.04$ for Lee and Curvature, $\tau = 0$ for Quad and Cubic, $\tau = 0.1$ for Cate 1 and 2, and $\tau = -3.35$ for Ludwig. Also displayed in Figure \ref{fig:meanFunctions} are a set of sample points $\{ (x_i, y_i), i = 1, \dots, 500 \}$ for each mean function, which shows what one dataset looks like for each mean function. One can see that---although $\epsilon_i \sim N(0, 0.1295^2)$ for all mean functions---the relative noise varies across mean functions.

For standard LLR and robust LLR we used the \texttt{rdrobust} \texttt{R} package \citep{calonico2017rdrobust}. For both methods, we used an MSE-optimal bandwidth that is the default in \texttt{rdrobust}. Simulation results using the bandwidth introduced in \cite{imbens2012optimal}---also known as the IK bandwidth, which has been widely used in practice---are provided in the Appendix in Table \ref{tab:simulationTableIK}, and results using the coverage error rate optimal bandwidth---an alternative bandwidth choice within the \texttt{rdrobust} \texttt{R} package that is also discussed in \cite{calonico2018effect}---are provided in the Appendix in Table \ref{tab:simulationTableCER}. The results using those bandwidths are largely the same as the results presented here. Simulation results for other bandwidth choices appear in \cite{calonico2014robust}. 

\begin{figure}[H]
  \centering
  \includegraphics[scale=0.7]{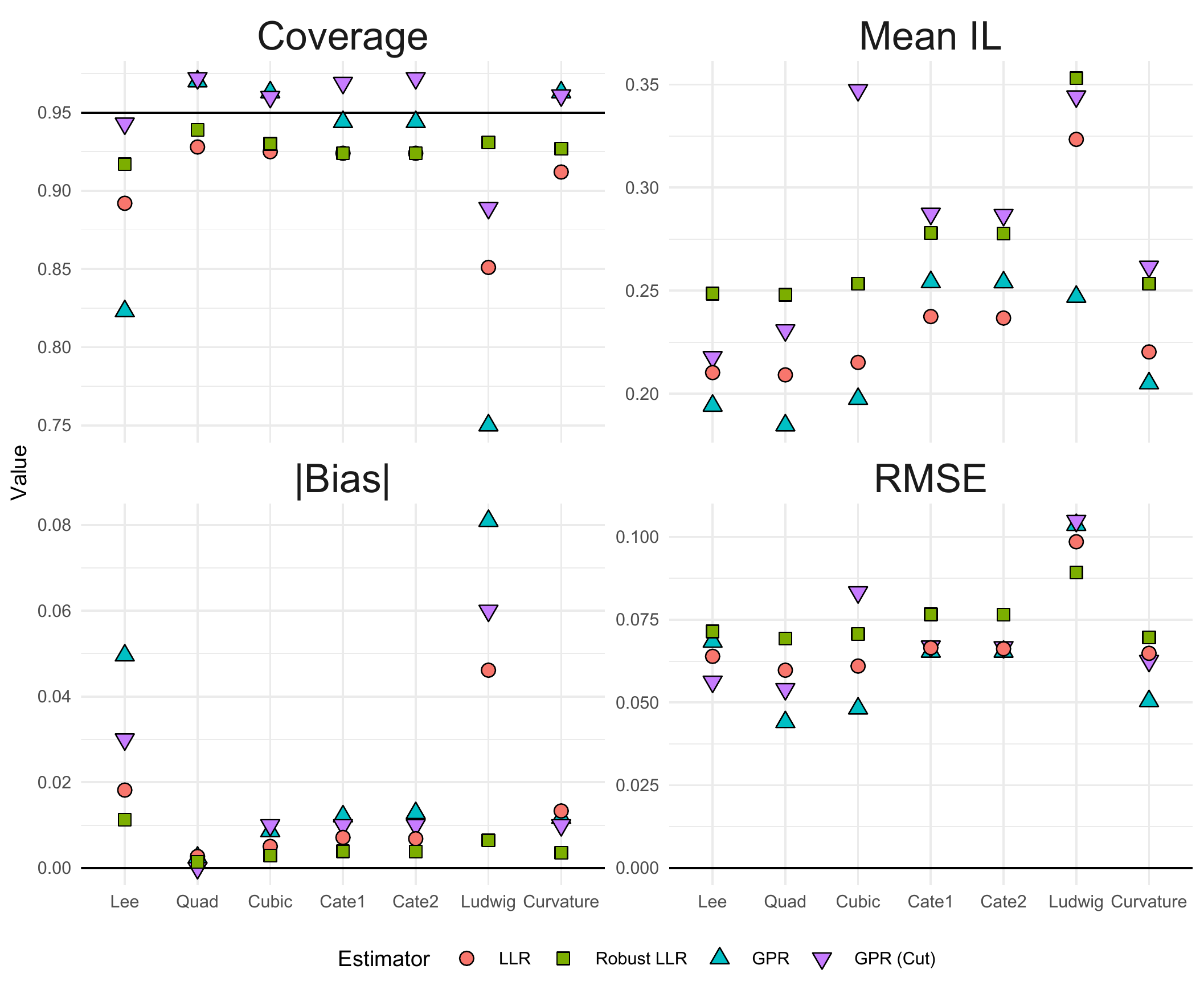}
  \caption{The coverage, mean interval length (IL), absolute bias, and root mean squared error (RMSE) for LLR, robust LLR, and our GPR method.}
  \label{fig:simSummary}
\end{figure}

\cite{imbens2012optimal} ran a simulation study that focused on the Lee, Quad, Cate 1, and Cate 2 mean functions, and they compared different bandwidth selectors for LLR in terms of bias and root mean squared error (RMSE). Similarly, \cite{calonico2014robust} ran a simulation study that focused on the Lee, Ludwig, and Curvature mean functions, and they compared different bandwidth selectors for LLR and their methodology in terms of coverage and mean interval length (IL). We synthesize these simulation studies and report in Figure \ref{fig:simSummary} how LLR, robust LLR, and two versions of our GPR methodology perform on the seven mean functions in Figure \ref{fig:meanFunctions} in terms of coverage, IL, absolute bias, and RMSE. The numbers plotted in Figure \ref{fig:simSummary} are in Tables \ref{tab:simulationTable} and \ref{tab:closeToBoundarySims} in the Appendix. Point estimates and 95\% confidence intervals for LLR and robust LLR were obtained from \texttt{rdrobust}. Point estimates and 95\% credible intervals for our methodology corresponded to the mean and 2.5\% and 97.5\% quantiles, respectively, of the posterior of the average treatment effect, shown in (\ref{posteriorTreatmentEffect}).

Robust LLR is meant to improve the coverage of LLR, and indeed it does for all datasets. The better coverage is in part due to wider confidence intervals (see the systematically higher mean interval length at the top right of Figure \ref{fig:simSummary}) and in part due to better bias properties (see the bottom left of Figure \ref{fig:simSummary}). Robust LLR also tends to exhibit worse RMSE than LLR (see the bottom right of Figure \ref{fig:simSummary}).

Our primary method (``GPR'') tends to exhibit narrower intervals than both LLR and robust LLR. GPR also exhibits better coverage than both methods, except for the Lee and Ludwig datasets. Furthermore, our method tends to exhibit lower RMSE than both LLR and robust LLR. However, our method always exhibits more bias than robust LLR, which explicitly uses a bias correction. 

\begin{figure}[H]
  \hspace{-0.5 in}\includegraphics[scale=0.3]{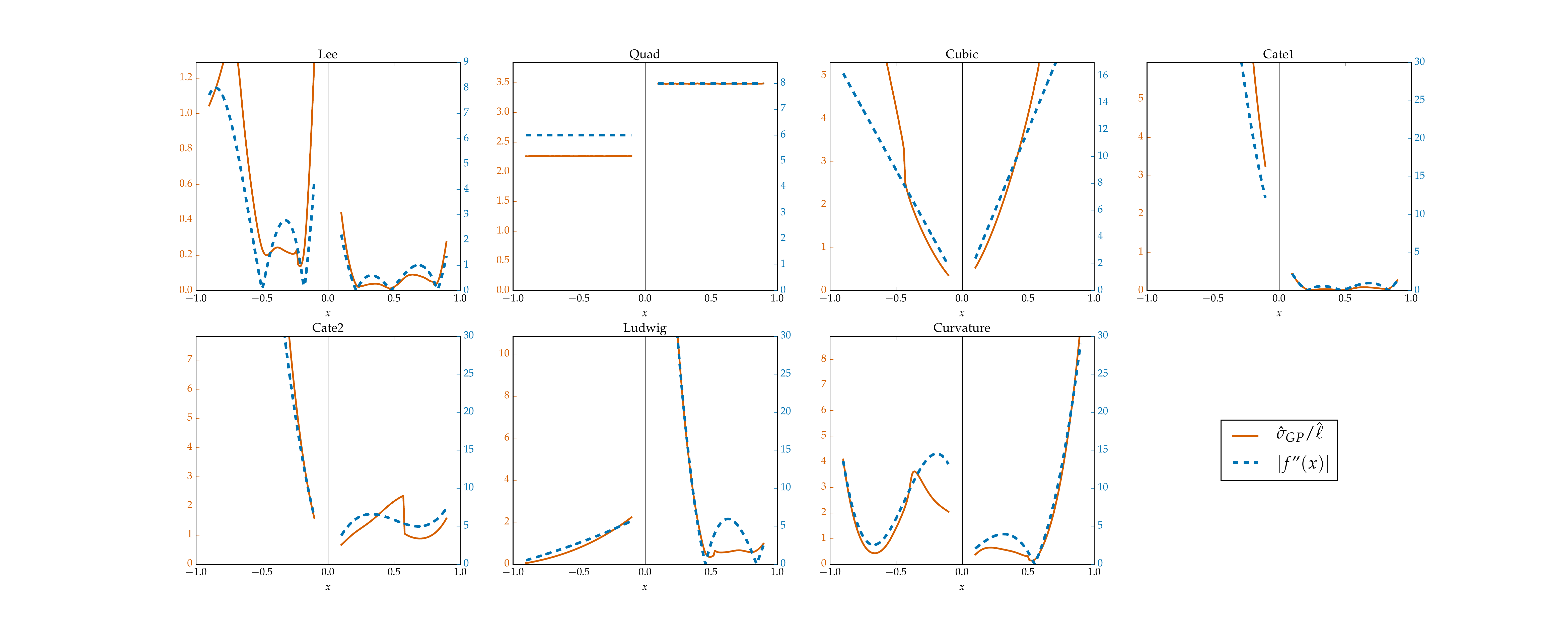}
  \caption{The absolute second derivative (blue) of the seven mean functions shown in Figure \ref{fig:meanFunctions}, and the ratio of the maximum-likelihood estimates of the variance and lengthscale (orange), computed within a sliding window of a noiseless version of the seven mean functions. The sliding window was the range $[x - 0.1, x + 1]$ for $x \in [-0.9, -0.1]$ in the control group (left-hand side) and $x \in [0.1, 0.9]$ in the treatment group (right-hand side).}
  \label{fig:secondDerivatives}
\end{figure}

For the Lee and Ludwig datasets, our method did worse than robust LLR in terms of coverage and bias, and this may be because it is inappropriate to assume that $\mu_T(x)$ and $\mu_C(x)$ are stationary processes---i.e., that the covariance parameters do not vary across the running variable $X$---in these cases. By assuming stationarity, our GPR model uses data both close to and far from the boundary to estimate the single variance $\sigma^2_{GP}$ and lengthscale $\ell$. This assumption is related to the stability of the second derivative of $\mu_T(x)$ and $\mu_C(x)$, because the covariance parameters of a Gaussian process dictate their derivative processes; see \cite{wang2012bayesian} for a further discussion of this relationship. Figure \ref{fig:secondDerivatives} displays the absolute second derivative of the seven mean functions (in blue). The Lee and Ludwig datasets are characterized by the absolute second derivative rapidly increasing near the boundary. Our GPR methodology likely does not do well for these datasets because we are using data far from the boundary to estimate the overall curvature of the mean function, which leads us to underestimating the curvature of the Lee and Ludwig mean functions at the boundary. 

Figure \ref{fig:secondDerivatives} also displays $\frac{\hat{\sigma}_{GP}}{\hat{\ell}}$, the ratio of the maximum-likelihood estimates of the covariance parameters, which was computed within a sliding window of a noiseless version of the seven mean functions. Although there is not a one-to-one correspondence between the absolute second derivative and $\frac{\hat{\sigma}_{GP}}{\hat{\ell}}$, their behavior is notably similar, which reinforces the idea that both the variance $\sigma^2_{GP}$ and lengthscale $\ell$ play a role in capturing the curvature of the mean function. Furthermore, this connection between the second derivative and the covariance parameters further suggests a similarity between the covariance parameters in our GPR methodology and the bandwidth in the LLR methodology, because the IK bandwidth is estimated as a nonlinear function of the estimated second derivative at the boundary $b$ \citep*{imbens2012optimal}.

If one did not believe the Stationary Assumption held, one alternative would be to only use data close to the boundary when fitting our GPR model. This is our second method, ``GPR (Cut),'' whose coverage, mean IL, absolute bias, and RMSE is also displayed in Figure \ref{fig:simSummary}. For each of the 1,000 replications of the seven mean functions, we first estimated the IK bandwidth with a rectangular kernel; then, we fit our GPR model within this estimated bandwidth. This procedure improves upon our GPR model for and only for the Lee and Ludwig datasets---the coverage increased to 94.3\% and 88.9\%, respectively, and the bias improved for the Lee and Ludwig datasets while staying the same for the other datasets. While the coverage also improved for the Cate 1 and Cate 2 datasets, this is likely due to the increase in the interval length. Because results improved only for these two datasets, this further demonstrates that using our GPR model on the whole dataset can be beneficial when $\mu_T(x)$ and $\mu_C(x)$ are stationary processes; otherwise, it may be preferable to only fit our GPR model to data close to the boundary.

Furthermore, this suggests that our method can be combined with a local randomization perspective for RDDs (e.g., \citealt{li2015evaluating}): One can first determine the window around the boundary where units are ``as-if randomized'' by using covariate balance tests such as \cite{cattaneo2015randomization} and \cite{li2015evaluating} and then use our GPR methodology within this window around the boundary. This approach is similar to \cite{cattaneo2017comparing}, who combined the local randomization perspective with parametric models for estimating the average treatment effect at the boundary.

Overall, GPR performs well compared to LLR and robust LLR. In particular, our GPR method tends to yield better interval length and RMSE properties than LLR and robust LLR, and it also yields better coverage when the underlying mean functions are stationary across the running variable $X$. The issue of undercoverage in LLR methodologies has been relatively unaddressed in the RDD literature, except for robust LLR \citep*{calonico2014robust}, and so our GPR method can be viewed as the second method to yield promising coverage properties for RDD analyses while also providing a flexible fit to the underlying mean functions $\mu_T(x)$ and $\mu_C(x)$. When $\mu_T(x)$ and $\mu_C(x)$ are nonstationary across $X$, our GPR methodology could be extended to incorporate a lengthscale function $\ell(x)$ instead of a single lengthscale $\ell$. However, such a lengthscale function would either need to be specified (see Rasmussen and Williams 2006, Chapter 4, for an example), or estimated via another Gaussian process prior \citep*{plagemann2008nonstationary}. In a similar vein, there has also been work on using dimension expansion to model nonstationary processes \citep{bornn2012modeling}. However, in an RDD, we only need a good estimate of the covariance parameters near the boundary, rather than across the entire mean functions $\mu_T(x)$ and $\mu_C(x)$. More work needs to be done to determine the optimal amount of data to include in our GPR model for estimating these parameters and the average treatment effect at the boundary in the case of nonstationary processes.

Now we compare how LLR, robust LLR, and GPR perform on a real dataset from the National Basketball Association.

\section{Empirical Example: The NBA Draft} \label{s:empiricalExample}

The National Basketball Association (NBA) draft, held annually, is divided into $2$ rounds, where each NBA teams gets one selection per round to draft a player of their choice. Because players are picked sequentially, there is no reason to believe there is a marked skill difference between the last pick of the first round and first pick of the second round. However, because of the difference in the perceived value of first-round versus second-round picks, as well as differing contract structures between the two rounds, we suspect that first-round picks are treated more favorably and given more playing time than their second-round colleagues, above and beyond what can be explained by differences in skill. As such, we seek to explore if there is a difference between first- and second-round picks in both skill and playing time.

\begin{figure}[H]
  \centering
  \includegraphics[scale=0.9]{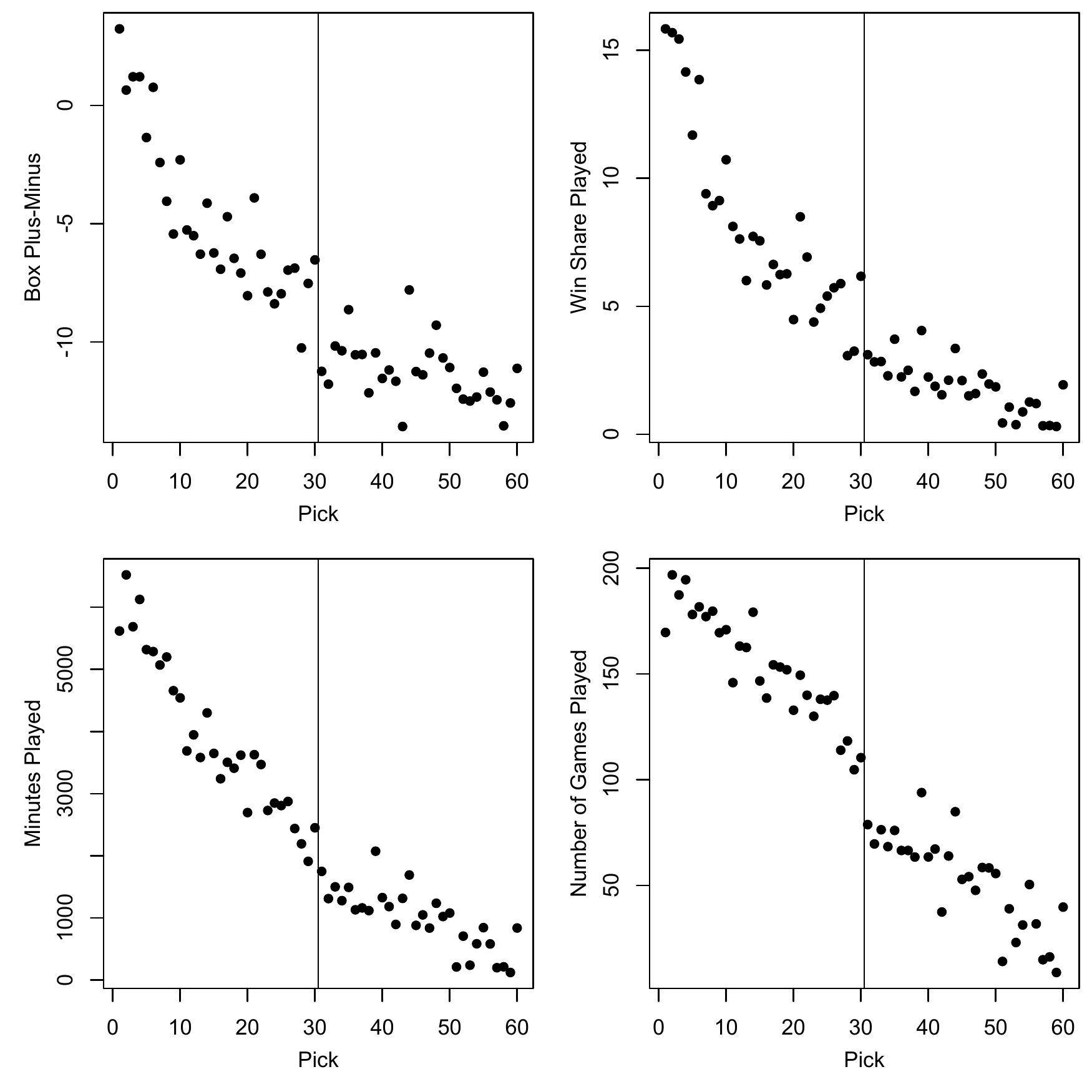}
\caption{Four basketball-player outcomes---number of minutes played, box plus-minus, win shares played, and number of games played---across picks.}
\label{fig:draftDataPlots}
\end{figure}

We want to estimate the treatment effect of having a second-round contract instead of a first-round contract on four basketball player outcomes: box plus-minus, win shares played, number of minutes played, and number of games played. The first two are overall measures of player performance \citep{kubatko2009calculating,myers2015box}, while the latter two are measures of playing time. Our data include the pick number and the four aforementioned outcomes for 1,238 NBA basketball players drafted between 1995 and 2016. Due to anomalies created by the NBA expanding from 29 teams to 30 teams in 2004, as well as some years with teams forfeiting picks, we shifted the pick numbers to ensure that $b = 30.5$ marked the discontinuity between the first- and second-round picks in each year. Figure \ref{fig:draftDataPlots} displays the NBA player data grouped by pick number for each of the four outcomes (after aligning pick numbers). Grouping by pick number allows us to understand the average performance and playing time of players drafted at each pick number, which is a standard approach in draft evaluation \citep{nbaDraft}.

Even though the running variable in this case is discrete---which can cause complications in some regression discontinuity analyses \citep{lee2008regression,kolesar2018inference}---we nonetheless apply LLR, robust LLR, and our GPR method to these data to compare how they perform in practice. As in Section \ref{s:simulations}, we used the \texttt{rdrobust} \texttt{R} package to implement LLR and robust LLR using the default MSE-optimal bandwidth. For GPR, we used the squared exponential covariance function assuming the Same Covariance Assumption; furthermore, we took the full-Bayesian approach to our GPR methodology and---similar to Section \ref{s:simulations}---placed independent $\mathcal{N}(0,100^2)$ priors on the mean function parameters and half-Cauchy$(0,5)$ priors on the covariance parameters.

\begin{figure}[H]
  \centering
  \includegraphics[scale=0.9]{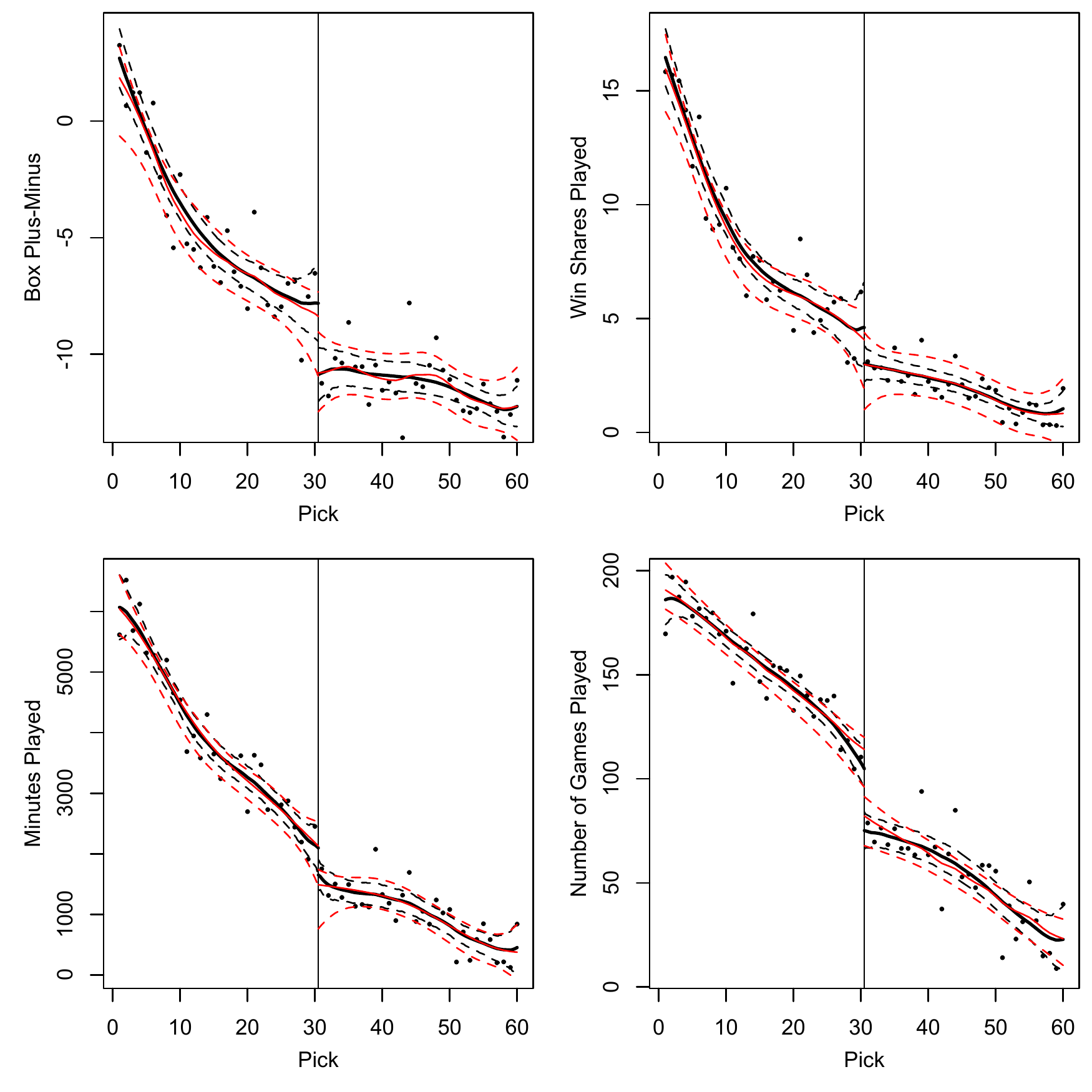}
  \caption{The estimated mean functions (solid lines) and corresponding confidence intervals (dashed lines) for LLR (black lines) and GPR (red lines). The lines for LLR were produced by the \texttt{rdd} R package \citep{dimmery2013rdd}, but using the bandwidth estimated by the \texttt{rdrobust} R package. The two treatment groups are the first round of picks (picks 30 and below) and second round of picks (picks 31 and above). We set the boundary to be $b = 30.5$ to minimize the amount of extrapolation that needs to be conducted on both sides of the boundary to estimate the treatment effect.}
  \label{fig:empiricalExamplePlot}
\end{figure}

Figure \ref{fig:empiricalExamplePlot} shows the estimated mean functions and corresponding confidence intervals for LLR and GPR, and Table \ref{tab:empiricalExampleAggregatedTable} shows the treatment effect point estimates and corresponding confidence intervals for LLR, robust LLR, and GPR.\footnote{Robust LLR yields an inflated confidence interval for the treatment effect specifically, which depends on a bias correction at the boundary. Thus, robust LLR does not yield confidence intervals for the entire mean functions, because the bias correction is boundary-specific. Thus, only LLR and GPR are displayed in Figure \ref{fig:empiricalExamplePlot}, while LLR, robust LLR, and GPR are all discussed in Table \ref{tab:empiricalExampleAggregatedTable}. Furthermore, for robust LLR in Table \ref{tab:empiricalExampleAggregatedTable}, we report the bias-corrected point estimate given by the \texttt{rdrobust} package.} The estimated mean functions for LLR and GPR are quite similar to each other for all outcomes, with GPR yielding slightly wider confidence intervals, as expected. All three methods find the number of games played for second-round picks to be significantly lower than that of first-round picks. Furthermore, LLR and robust LLR find the box plus-minus for second-round picks to be significantly lower than that of first-round picks; meanwhile, GPR finds this difference to be borderline insigificant. Out of these three methods, the results from our GPR method are most in line with previous reports---both qualitative \citep{huffPost} and quantitative \citep{harvardSports}---on the difference between first- and second-round NBA basketball players, which have claimed that there is a difference in attention given to first-round picks (e.g., games played) but not a difference in player ability (e.g., box plus-minus and win shares).

 \begin{table}
{\centering
\captionof{table}{Treatment effect estimation for LLR, robust LLR, and GPR on NBA data} \label{tab:empiricalExampleAggregatedTable}
\noindent\resizebox{\linewidth}{!}{%
  \begin{tabular}{|c|| c |c|| c |c||  c |c|}
  \hline
  \cline{2-7}
  & \multicolumn{2}{c||}{\textbf{LLR}} & \multicolumn{2}{c||}{\textbf{Robust LLR}} & \multicolumn{2}{c|}{\textbf{GPR}} \\
  \hline
    \textbf{Outcome} & Estimate & 95\% CI & Estimate & 95\% CI & Estimate & 95\% CI \\
  \hline
  \multicolumn{1}{|l||}{Box Plus-Minus} & \textbf{-3.06}  & [-5.20, -0.92]  & \textbf{-3.37}  & [-5.86, -0.88]   & -2.42 & [-5.02, 0.01]  \\
     \hline
  \multicolumn{1}{|l||}{Win Shares} & -1.58  & [-4.09, 0.93]  & -1.73  & [-4.76, 1.29]   & -1.08  & [-3.04, 0.88] \\
   \hline 
     \multicolumn{1}{|l||}{Minutes Played} & -446.29 & [-1042.81, 150.23]   & -406.63  & [-1123.99, 310.72]  &  -640.75 & [-1259.65, 5.51]  \\
  \hline
  \multicolumn{1}{|l||}{Games Played} & \textbf{-29.71}  & [-40.20, -19.22]  & \textbf{-28.16}  & [-40.81, -15.51] & \textbf{-32.00}  & [-45.69, -18.14]  \\
     \hline 
  \end{tabular} 
  }%
  \par
  }
  \bigskip
  Point estimates and 95\% confidence intervals for the treatment effect on each of the four outcomes: box plus-minus, win shares played, number of minutes played, and number of games played. Statistically significant point estimates are in bold.
\end{table}

In summary, using our GPR methodology, we find that the treatment effect of being a second-round pick significantly reduces the number of games played and marginally reduces the number of minutes played. This suggests that there is a drop-off in playing time for second-round players relative to their first-round counterparts beyond that explainable by the natural drop-off in playing time between successive picks. Furthermore, we find that there is not a significant difference in player ability between first- and second-round basketball players at the boundary between the 30th and 31st picks that divides the first and second rounds of the NBA draft.

\section{Discussion and Conclusion} \label{s:conclusion}

Local linear regression (LLR) and its variants are currently the most common methodologies for estimating the average treatment effect at the boundary in RDDs. These methods are popular because they are easy to implement and there is a large literature on their theoretical properties. However, recent works have noted that LLR tends to yield confidence intervals that undercover, and new methodologies---namely that of \cite{calonico2014robust}---have tried to address this issue. As an alternative to LLR, we proposed a Gaussian Process regression (GPR) methodology that flexibly fits the treatment and control response functions by placing a general prior on the mean response functions. We showed via simulation that our GPR methodology tends to outperform standard LLR and the state-of-the-art methodology of \cite{calonico2014robust} in terms of coverage, interval length, and RMSE. Furthermore, we used our GPR methodology on a real-world sharp RDD in the National Basketball Association (NBA) draft and found that GPR yielded results that were more in line with previous reports on the NBA draft than were results from LLR methods. Overall, our methodology addresses the undercoverage issue commonly seen in RDDs without sacrificing too much power to detect treatment effects, thereby adding to the growing literatures on improving inference for RDDs \citep{calonico2014robust,calonico2018optimal} and on Bayesian methods for RDDs \citep{chib2014nonparametric,chib2015,geneletti,li2015evaluating}.

Our methodology focuses on flexibly fitting the mean treatment and control responses while also improving coverage properties; however, there are many other issues of interest in the RDD literature. For example, we only consider sharp RDDs; \cite{li2015evaluating} and \cite{chib2015} provide Bayesian methodologies for fuzzy RDDs which could likely be combined with our Gaussian process approach. Furthermore, we focused on RDDs that only have one background covariate---the running variable---but other covariates could be included in our GPR methodology to improve the precision of our treatment effect estimator (e.g., \citealt{calonico2016regression}). See \cite{imbens2008regression} for a review of these other RDD concerns.

Furthermore, while our method can incorporate prior knowledge and uncertainty in various parameters that are typically discussed in the RDD literature, it is not necessarily clear when this should be done for our GPR model. For example, \cite{hall2001bootstrapping} found that incorporating the uncertainty in the bandwidth---which is somewhat analogous to the covariance parameters in our GPR model---for density estimators via bootstrapping can be inconsequential or even detrimental in some cases. Although our simulation study suggests that it is beneficial to take a full-Bayesian approach and propogate uncertainty in the GPR model parameters when estimating treatment effects in RDDs, more work needs to be done to determine when it is most appropriate to incorporate uncertainty in these parameters. Furthermore, we only focused on the squared exponential covariance function for Gaussian processes, but other covariance functions should be considered. Because estimating the average treatment effect at the boundary in an RDD is fundamentally an extrapolation issue, covariance functions whose purpose is to extrapolate well (e.g., \citealt{wilson2013gaussian}) may be particularly suitable for RDDs.  

Additionally, although our GPR methodology exhibits arguably better coverage, interval length, and RMSE properties than standard methodologies in the literature, there are ways our methodology could be improved, even for the case of a sharp RDD with only one covariate. Our methodology does not perform well when the mean response functions $\mu_T(x)$ and $\mu_C(x)$ are nonstationary across the running variable $X$. More work needs to be done to model nonstationary processes within the context of RDDs, such as by estimating a lengthscale function $\ell(x)$ or by determining the optimal amount of data to use when estimating the covariance parameters for our GPR methodology. Relatedly, our simulation results suggest that our GPR methodology can be used in combination with a local randomization framework for RDDs (such as that seen in \citealt{li2015evaluating}).

Finally, a promising avenue for future research is extending our GPR methodology beyond one-dimensional sharp RDDs. In particular, recent work has explored geographic RDDs, where spatial variation in outcomes must be accounted for when estimating the treatment effect along a geographic boundary \citep{keele2014geographic,keele2015enhancing}. Because GPR has been widely used in spatial statistics \citep{banerjee2014hierarchical,cressie2015statistics}, it may be particularly suitable for geographic RDDs. We explore the use of our GPR methodology for geographic RDDs in \cite{rischard2018bayesian}.

\section{Appendix}

\subsection{Complete Simulation Results from Section \ref{s:simulations}}

 \begin{table}[H]
{\centering
\captionof{table}{Simulation for $n = 500$, shown in Figure \ref{fig:simSummary}} \label{tab:simulationTable}
\noindent\resizebox{1.1\linewidth}{!}{%
  \hskip-2.cm\begin{tabular}{|c|c c c c||c c c c|| c c c c|}
  \hline
  & \multicolumn{12}{c|}{Method} \\
  \cline{2-13}
  & \multicolumn{4}{c||}{\textbf{LLR}} & \multicolumn{4}{c||}{\textbf{Robust LLR}} & \multicolumn{4}{c|}{\textbf{GPR}} \\
  \hline
    Dataset & \textbf{EC} & \textbf{$\overline{\text{IL}}$} & \textbf{Bias} & \textbf{RMSE} & \textbf{EC} & \textbf{$\overline{\text{IL}}$} & \textbf{Bias} & \textbf{RMSE} & \textbf{EC} & \textbf{$\overline{\text{IL}}$} & \textbf{Bias} & \textbf{RMSE} \\ 
  \hline
  \multicolumn{1}{|l|}{Lee} & 89.2\%  & 0.21 & 0.02  & 0.06  & 91.7\% & 0.25 & 0.01  & 0.07 & 82.3\% & 0.19 & 0.05 & 0.07 \\ 
  \hline
  \multicolumn{1}{|l|}{Quad} & 92.8\%  & 0.21 & 0.00   & 0.06  & 93.9\% & 0.25 &  0.00  & 0.07  & 97.0\% & 0.18 & 0.00 & 0.04 \\
     \hline
  \multicolumn{1}{|l|}{Cubic} & 92.5\%  & 0.22 & -0.01 & 0.06  & 93.0\% & 0.25 & 0.00 & 0.07 & 96.3\% & 0.20 & -0.01 & 0.05 \\
  \hline
  \multicolumn{1}{|l|}{Cate 1} & 92.4\%  & 0.24 &  -0.01 & 0.07  & 92.4\% & 0.28 &  0.00  &  0.08  & 94.4\% & 0.25 &  -0.01 & 0.07 \\
     \hline
  \multicolumn{1}{|l|}{Cate 2} & 92.4\%  & 0.24 & -0.01  & 0.07   & 92.4\% & 0.28 & 0.00   &  0.08  & 94.4\% & 0.25 & -0.01 & 0.07 \\
     \hline
  \multicolumn{1}{|l|}{Ludwig} & 85.1\%  & 0.32 & 0.05  & 0.10   & 93.1\% & 0.35 & 0.01   &  0.09  & 75.0\% & 0.25 & 0.08 & 0.10 \\
     \hline
  \multicolumn{1}{|l|}{Curvature} & 91.2\%  & 0.22 & -0.01   & 0.06   & 92.7\% & 0.25 & 0.00   &  0.07  & 96.3\% & 0.21 & -0.01 & 0.05 \\
     \hline    
  \end{tabular} 
  }%
  \par
  }
  \bigskip
  Simulations assessing the empirical coverage (EC), mean interval length ($\overline{\text{IL}}$), bias, and root mean squared error (RMSE) for local linear regression (LLR), robust LLR, and our Gaussian Process Regression (GPR) method, where we used the MSE-opitmal default bandwidth in \texttt{rdrobust} when implementing LLR and robust LLR. These methods were performed on 1,000 replications of seven different datasets, which were also used in \cite{imbens2012optimal} and \cite{calonico2014robust}. A plot of these numbers is shown in Figure \ref{fig:simSummary}.
\end{table}

 \begin{table}[H]
{\centering
\captionof{table}{Simulation Results for GPR, GPR (Cut), GPR (Zero Mean), and GPR (MLE)} \label{tab:closeToBoundarySims}
\resizebox{1\linewidth}{!}{%
  \begin{tabular}{|c|c c c c|c c c c|c c c c|c c c c|}
  \hline
  & \multicolumn{16}{c|}{\textbf{Method}} \\
  \cline{2-17}
  & \multicolumn{4}{c|}{\textbf{GPR}} & \multicolumn{4}{c|}{\textbf{GPR (Cut)}} & \multicolumn{4}{c|}{\textbf{GPR (Zero Mean)}} & \multicolumn{4}{c|}{\textbf{GPR (MLE)}} \\
  \hline
    \textbf{Data} & \textbf{EC} & \textbf{$\overline{\text{IL}}$} & \textbf{Bias} & \textbf{RMSE} & \textbf{EC} & \textbf{$\overline{\text{IL}}$} & \textbf{Bias} & \textbf{RMSE} & \textbf{EC} & \textbf{$\overline{\text{IL}}$} & \textbf{Bias} & \textbf{RMSE} & \textbf{EC} & \textbf{$\overline{\text{IL}}$} & \textbf{Bias} & \textbf{RMSE} \\
  \hline
  \multicolumn{1}{|l|}{Lee} & 82.3\% & 0.19 & 0.05 & 0.07 & 94.3\% & 0.22 & 0.03 & 0.06 & 81.6\% & 0.18 & 0.05 & 0.07 & 76.4\% &  0.15 & 0.04 & 0.06 \\ 
  \hline
  \multicolumn{1}{|l|}{Quad} & 97.0\% & 0.18 & 0.00 & 0.04 & 97.2\% &  0.23 & -0.00 & 0.05 & 97.0\% & 0.19 & 0.01 & 0.04 & 95.9\% &  0.18 &  0.00 &  0.05 \\
     \hline
  \multicolumn{1}{|l|}{Cubic} & 96.3\% & 0.20 & -0.01 & 0.05 & 96.0\% & 0.35 & -0.01 &  0.08 & 96.7\% & 0.20 & 0.00 & 0.05 & 91.1\% &  0.19 & -0.03 &  0.06 \\
  \hline
  \multicolumn{1}{|l|}{Cate 1} & 94.4\% & 0.25 &  -0.01 & 0.07 & 96.9\% & 0.29 & -0.01 & 0.07 & 95.2\% & 0.26 & 0.00 & 0.07 & 93.8\% &  0.25 &  0.00 &  0.07 \\
     \hline
  \multicolumn{1}{|l|}{Cate 2} & 94.4\% & 0.25 & -0.01 & 0.07 & 97.2\% &  0.29 & -0.01 & 0.07 & 95.4\% & 0.26 & 0.00 & 0.07 & 94.2\% &  0.25 &  0.00 &  0.07 \\
     \hline
  \multicolumn{1}{|l|}{Ludwig} & 75.0\% & 0.25 & 0.08 & 0.10 & 88.9\% &  0.34 & 0.06 & 0.10 & 64.5\% & 0.24 & 0.10 & 0.12 & 52.0\% &  0.24 &  0.11 &  0.13 \\
     \hline
  \multicolumn{1}{|l|}{Curvature} & 96.3\% & 0.21 & -0.01 & 0.05 & 96.1\% & 0.26 & -0.01 &  0.06 & 96.9\% & 0.20 & 0.00 & 0.05 & 89.6\% &  0.20 & -0.03 &  0.06 \\
     \hline    
  \end{tabular} 
  }%
  \par
  }
  \bigskip
  Simulation results assessing the empirical coverage (EC), mean interval length ($\overline{\text{IL}}$), bias, and root mean squared error (RMSE) for GPR (which uses the full Bayesian approach discussed in Section \ref{ss:covarianceUncertainty}), GPR using only data within the IK bandwidth with a rectangular kernel (called GPR (Cut)), GPR using a zero mean function in the prior (\ref{linearMeanFunction}) instead of a linear trend (called GPR (Zero Mean)), and GPR plugging in the MLE for the covariance parameters (calld GPR (MLE)). Note that the GPR columns are the same as those in Table \ref{tab:simulationTable}. GPR (Cut) performs much better than GPR for the Lee and Ludwig datasets, and GPR (Cut) performs marginally better than GPR for the Cate 1 and Cate 2 datasets. Otherwise, GPR (Cut) is equal to or worse than GPR. This demonstrates that GPR on the whole dataset can be beneficial when $\mu_T(x)$ and $\mu_C(x)$ are stationary processes; otherwise, it may be preferable to only fit our GPR model to data close to the boundary. Furthermore, GPR (Zero Mean) performs similarly to GPR for most of the datasets; this suggests that our results are generally insensitive to specification of the mean function in the Gaussian process prior. However, GPR does perform better than GPR (Zero Mean) for the Ludwig dataset. This is likely because, as discussed in Section \ref{s:simulations}, when GPR extrapolates to the boundary, its predictions will be pulled towards the global linear trend, while predictions from GPR (Zero Mean) will be pulled towards the global mean. This also suggests why, for the Ludwig dataset, the bias for GPR (Zero Mean) is higher than the bias for GPR. Finally, GPR that simply plugs in the MLE of the covariance parameters tends to perform worse than the full-Bayesian GPR approach, especially in terms of coverage. This further suggests that incorporating uncertainty in the covariance parameters for our GPR method can lead to promising inferential properties. 
\end{table}

 \begin{table}[H]
{\centering
\captionof{table}{Simulation for $n = 500$, using the IK bandwidth} \label{tab:simulationTableIK}
\noindent\resizebox{1.1\linewidth}{!}{%
  \hskip-2.cm\begin{tabular}{|c|c c c c||c c c c|| c c c c|}
  \hline
  & \multicolumn{12}{c|}{Method} \\
  \cline{2-13}
  & \multicolumn{4}{c||}{\textbf{LLR}} & \multicolumn{4}{c||}{\textbf{Robust LLR}} & \multicolumn{4}{c|}{\textbf{GPR}} \\
  \hline
    Dataset & \textbf{EC} & \textbf{$\overline{\text{IL}}$} & \textbf{Bias} & \textbf{RMSE} & \textbf{EC} & \textbf{$\overline{\text{IL}}$} & \textbf{Bias} & \textbf{RMSE} & \textbf{EC} & \textbf{$\overline{\text{IL}}$} & \textbf{Bias} & \textbf{RMSE} \\
  \hline
  \multicolumn{1}{|l|}{Lee} & 82.7\%  & 0.15 & 0.04  & 0.05  & 91.2\% & 0.27 & 0.01  & 0.08 & 82.3\% & 0.19 & 0.05 & 0.07 \\ 
  \hline
  \multicolumn{1}{|l|}{Quad} & 94.6\%  & 0.15 & 0.00   & 0.04  & 93.2\% & 0.24 &  0.00  & 0.07  & 97.0\% & 0.18 & 0.00 & 0.04 \\
     \hline
  \multicolumn{1}{|l|}{Cubic} & 93.7\%  & 0.19 & -0.01 & 0.05  & 93.8\% & 0.24 & 0.01 & 0.06 & 96.3\% & 0.20 & -0.01 & 0.05 \\
  \hline
  \multicolumn{1}{|l|}{Cate 1} & 91.2\%  & 0.21 &  -0.01 & 0.06  & 92.9\% & 0.26 &  0.01  &  0.07  & 94.4\% & 0.25 &  -0.01 & 0.07 \\
     \hline
  \multicolumn{1}{|l|}{Cate 2} & 92.1\%  & 0.22 & -0.01  & 0.06   & 92.8\% & 0.26 & 0.01   &  0.07  & 94.4\% & 0.25 & -0.01 & 0.07 \\
     \hline
  \multicolumn{1}{|l|}{Ludwig} & 31.5\%  & 0.23 & 0.15  & 0.16   & 89.8\% & 0.27 & 0.04   &  0.08  & 75.0\% & 0.25 & 0.08 & 0.10 \\
     \hline
  \multicolumn{1}{|l|}{Curvature} & 84.4\%  & 0.19 & -0.03   & 0.06   & 94.8\% & 0.23 & 0.00   &  0.06  & 96.3\% & 0.21 & -0.01 & 0.05 \\
     \hline    
  \end{tabular} 
  }%
  \par
  }
  \bigskip
  Simulations assessing the empirical coverage (EC), mean interval length ($\overline{\text{IL}}$), bias, and root mean squared error (RMSE) for local linear regression (LLR), robust LLR, and our Gaussian Process Regression (GPR) method, where we used the bandwidth of \cite{imbens2012optimal} when implementing LLR and robust LLR. These results are largely the same as the results presented in Table \ref{tab:simulationTable}: Our method performed best in terms of coverage, except for the Lee and Ludwig datasets. Furthermore, our method tended to yield narrower intervals and lower RMSE than robust LLR. Our method also tended to yield more bias than robust LLR.
\end{table}

 \begin{table}[H]
{\centering
\captionof{table}{Simulation for $n = 500$, using the CER bandwidth} \label{tab:simulationTableCER}
\noindent\resizebox{1.1\linewidth}{!}{%
  \hskip-2.cm\begin{tabular}{|c|c c c c||c c c c|| c c c c|}
  \hline
  & \multicolumn{12}{c|}{Method} \\
  \cline{2-13}
  & \multicolumn{4}{c||}{\textbf{LLR}} & \multicolumn{4}{c||}{\textbf{Robust LLR}} & \multicolumn{4}{c|}{\textbf{GPR}} \\
  \hline
    Dataset & \textbf{EC} & \textbf{$\overline{\text{IL}}$} & \textbf{Bias} & \textbf{RMSE} & \textbf{EC} & \textbf{$\overline{\text{IL}}$} & \textbf{Bias} & \textbf{RMSE} & \textbf{EC} & \textbf{$\overline{\text{IL}}$} & \textbf{Bias} & \textbf{RMSE} \\
  \hline
  \multicolumn{1}{|l|}{Lee} & 91.1\%  & 0.25  & 0.01  &  0.07 & 91.8\% & 0.27 & 0.01  & 0.08 & 82.3\% & 0.19 & 0.05 & 0.07 \\ 
  \hline
  \multicolumn{1}{|l|}{Quad} & 92.3\%  & 0.24  & -0.00    & 0.07  & 92.7\% & 0.27  & -0.00   & 0.08  & 97.0\% & 0.18 & 0.00 & 0.04 \\
     \hline
  \multicolumn{1}{|l|}{Cubic} & 92.2\%  & 0.25  & -0.00 & 0.07  & 92.1\% & 0.27  & 0.00 & 0.08 & 96.3\% & 0.20 & -0.01 & 0.05 \\
  \hline
  \multicolumn{1}{|l|}{Cate 1} & 91.9\%  & 0.28  & -0.00  &  0.08  & 92.3\% & 0.30  & 0.00   & 0.08   & 94.4\% & 0.25 &  -0.01 & 0.07 \\
     \hline
  \multicolumn{1}{|l|}{Cate 2} & 92.0\%  & 0.28  & -0.00   &  0.08  & 92.3\% & 0.30  & 0.00   & 0.08  & 94.4\% & 0.25 & -0.01 & 0.07 \\
     \hline
  \multicolumn{1}{|l|}{Ludwig} & 91.5\%  & 0.39  & 0.02   &  0.10  & 92.9\% & 0.41  & 0.00   & 0.10  & 75.0\% & 0.25 & 0.08 & 0.10 \\
     \hline
  \multicolumn{1}{|l|}{Curvature} & 91.4\%  & 0.26  & -0.00  &  0.07 & 93.3\% & 0.28  & 0.00   & 0.08  & 96.3\% & 0.21 & -0.01 & 0.05 \\
     \hline    
  \end{tabular} 
  }%
  \par
  }
  \bigskip
  Simulations assessing the empirical coverage (EC), mean interval length ($\overline{\text{IL}}$), bias, and root mean squared error (RMSE) for local linear regression (LLR), robust LLR, and our Gaussian Process Regression (GPR) method, where we used the coverage error rate (CER) optimal bandwidth---an alternative bandwidth choice within the \texttt{rdrobust} \texttt{R} package that is also discussed in \cite{calonico2018effect}. These results are largely the same as the results presented in Tables \ref{tab:simulationTable} and \ref{tab:simulationTableIK}: Our method performed best in terms of coverage, except for the Lee and Ludwig datasets. Furthermore, our method tended to yield narrower intervals and lower RMSE than robust LLR. Our method also tended to yield more bias than robust LLR.
\end{table}

\subsection{Assumptions for Posterior Consistency Proofs}

\textbf{Assumptions on the Running Variable $X$ (Assumptions A1)}
\begin{enumerate}
  \item Assume the control running variable values $\{x_i\}_{i=1}^{n_C}$ are known elements of $[b_C, b]$ and the treatment running variable values $\{x_i\}_{i=1}^{n_T}$ are known elements of $[b, b_T]$, for some $b_C, b_T$, and boundary $b$.
\end{enumerate}
\textbf{Assumptions on the Response Function (Assumptions A2)}
\begin{enumerate}
  \item Let $C^{\alpha}[b, b_T]$ and $C^{\alpha}[b_C, b]$ be H$\ddot{\text{o}}$lder spaces of $\alpha$-smooth functions $f: [b, b_T] \rightarrow \mathbb{R}$ and $f: [b_C, b] \rightarrow \mathbb{R}$, respectively. Assume $\mu_T(x) \in C^{\alpha}[b, b_T]$ and $\mu_C(x) \in C^{\alpha}[b_C, b]$, where $\mu_T(x)$ and $\mu_C(x)$ are the treatment and control response functions.
  \item The treatment and control responses $\{y_i\}_{i=1}^{n_T}$ and $\{y_i\}_{i=1}^{n_C}$ have the following relationships with the running variable:
  \begin{align}
    y_i &= \mu_T(x_i) + \epsilon_i, \hspace{0.1 in} i = 1, \dots, n_T \\
    y_i &= \mu_C(x_i) + \epsilon_i, \hspace{0.1 in} i = 1, \dots, n_C
  \end{align}
  for mean response functions $\mu_T(x)$ and $\mu_C(x)$ and independent errors $\{\epsilon_i\}_{i=1}^{n_C} \sim N(0, \sigma^2_{y0})$ and $\{\epsilon_i\}_{i=1}^{n_T} \sim N(0, \sigma^2_{y1})$.
\end{enumerate}
\textbf{Assumptions on the Noise (Assumptions A3)}
\begin{enumerate}
  \item The priors on $\sigma_{y0}$ and $\sigma_{y1}$ have support on compact intervals that are subsets of $(0, \infty)$ which contain the true errors $\sigma_{y0}$ and $\sigma_{y1}$, respectively.
\end{enumerate}
\textbf{Assumptions on the Prior for $\ell$ (Assumptions A4)}

\begin{enumerate}
  \item The lengthscale $\ell$ has a prior distribution $\kappa$ such that, for positive constants $C_1, D_1, C_2, D_2$, nonnegative constants $p, q$, and every sufficiently large $a >0$,
\begin{align}
  C_1 a^p \exp(-D_1 a \log^q a) \leq \kappa(a) \leq C_2 a^p \exp(-D_2 a \log^q a)
\end{align}
\end{enumerate}

\subsection{Proof of Theorems 1 and 2}

\textbf{Proof of Theorem 1}: Assume that the Stationary Assumption holds, the covariance functions $K_T(x, x)$ and $K_C(x, x)$ are fixed, and Assumptions A1, A2, and A3 hold. Then, according to Theorem 2.2 in \cite{van2009adaptive}, for sufficiently large $M_T$ and $M_C$,
\begin{equation}
\begin{aligned}
  &\prod \left( \mu_T: h(\mu_T, \mu_T^*) \geq M_T \epsilon_{n_T} | x_1, \dots, x_{n_T} \right) \xrightarrow{P_{\mu_T^*}} 0, \hspace{0.1 in} \text{and} \\
  &\prod \left( \mu_C: h(\mu_C, \mu_C^*) \geq M_C \epsilon_{n_C} | x_1, \dots, x_{n_C} \right) \xrightarrow{P_{\mu_C^*}} 0 \label{separateConsistencies}
\end{aligned}
\end{equation}
for some contraction rates $\epsilon_{n_T}$ and $\epsilon_{n_C}$, where $h$ is the Hellinger distance. The nature of the contraction rates $\epsilon_{n_T}$ and $\epsilon_{n_C}$ are discussed in \cite{van2009adaptive}, and depend on the differentiability and smoothness of the true $\mu_T^*(x)$ and $\mu_C^*(x)$. Note that the only covariate value for which both of these hold is $x = b$, because by Assumption A1, the intersection of the supports of $\{x_i\}_{i=1}^{n_C}$ and $\{x_i\}_{i=1}^{n_T}$ is only the boundary $b$. 

Because the Hellinger distance is symmetric and satisfies the triangle inequality,
\begin{align}
  h(\tau, \tau^*) &= h \left(\mu_T(b) - \mu_C(b), \mu_T^*(b) - \mu_C^*(b) \right) \\
  &\leq h(\mu_T(b), \mu_T^*(b)) + h(\mu_C(b), \mu_C^*(b)) \\
  &\leq M_T \epsilon_{n_T} + M_C \epsilon_{n_C}
\end{align}
where the last inequality holds with posterior probability 1, by (\ref{separateConsistencies}). Therefore,
\begin{align}
  \prod \left( \tau: h(\tau, \tau^*) \geq M \epsilon_{n} | x_1, \dots, x_{n} \right) \xrightarrow{P_{\tau^*}} 0
\end{align}
where $M \equiv \sqrt{2} \cdot \text{max}(M_T, M_C)$ and $\epsilon_n \equiv \sqrt{2} \cdot \text{max}(\epsilon_{n_T}, \epsilon_{n_C})$, so that $M \epsilon_n \geq M_T \epsilon_{n_T} + M_C \epsilon_{n_C}$. $\blacksquare$ \\

\textbf{Proof of Theorem 2}: Assume that the Stationary Assumption holds, the $\sigma^2_{GP}$ parameters in $K_T(x, x)$ and $K_C(x, x)$ are fixed, and Assumptions A1, A2, A3, and A4 hold. By Theorem 3.1 of \cite{van2009adaptive}, (\ref{separateConsistencies}) holds, and then the proof of Theorem 2 is identical to that of Theorem 1. $\blacksquare$

\subsection{Extending Theorems 1 and 2 to Other Mean and Covariance Functions and Random Variance}

\citealt{rasmussenWilliams} (Section 2.7) discuss other choices of mean functions besides $m(x) = 0$, particularly mean functions of the form $m(x) = \mathbf{h}(x)^T \boldsymbol{\beta}$ for some fixed basis functions $\mathbf{h}(x)$. However, \cite{rasmussenWilliams} argue that different choices of $m(x)$ are more for interpretability than predictive accuracy, because $m(x) = 0$ does not constrain the posterior mean to be zero, and so different choices of $m(x)$ likely do not affect the consistency results in Section \ref{s:consistency}. To the best of our knowledge, the literature has focused primarily on the choice $m(x) = 0$ for posterior consistency of GPR (e.g., \citealt{choiAndSchervish} and \citealt{van2009adaptive}). 

\cite{van2009adaptive} discuss how their Theorems 2.2 and 3.1 (which correspond to our Theorems 1 and 2, respectively) extend to covariance functions besides the squared exponential. Specifically, their results hold for processes whose spectral measure has subexponential tails; the squared exponential process falls under this class, but \cite{van2009adaptive} discuss other processes that fall under this class as well, and our Theorems 1 and 2 also hold for those processes. 

Finally, the literature has focused on posterior consistency for the case when a prior is placed on the lengthscale $\ell$ but not on the variance $\sigma^2_{GP}$, as in \cite{van2009adaptive}. To the best of our knowledge, \cite{choi2007} is the only work to consider posterior consistency when priors are placed on both $\ell$ and $\sigma^2_{GP}$; however, these results only hold for binary Gaussian process regression, and their necessary assumptions are more restrictive than those presented in this paper. We leave posterior consistency of our GPR method when priors are placed on both $\ell$ and $\sigma^2_{GP}$ as future work.

\newpage

\bibliography{jspiGPRRDDSubmissionBib}

\bibliographystyle{apa-good}

\end{document}